\documentclass{aa}
\usepackage{tabularx}
\usepackage{amssymb}
\usepackage{amsmath}
\usepackage{txfonts}
\usepackage{balance}
\usepackage{natbib}
\bibpunct{(}{)}{;}{a}{}{,} 
\usepackage{color}
\usepackage{xspace}
\usepackage[normalem]{ulem} 
\usepackage[table]{xcolor} 
\usepackage{multirow}
\usepackage{rotating}
\usepackage{bbm}
\usepackage{cancel}
\usepackage{graphicx}
\usepackage{caption}
\usepackage{ifthen}
\usepackage{booktabs}
\usepackage{tabularx}
\usepackage{float}
\usepackage[caption=false]{subfig}
\newcommand{\e}[1]{\ensuremath{\times 10^{#1}}}

\makeatletter
\if@referee
    \newcommand{\com}[1]{\textcolor{red}{[#1]}}                            
    \usepackage[nolists,noheads,nomarkers]{endfloat}                       
\else
    \newcommand{\com}[1]{}                                                 
\fi
\makeatother
\usepackage[%
pdfauthor={Fredrik Windmarkl},
pdftitle={TBD},
pdfstartview=FitH,
linkcolor=blue,
anchorcolor=blue,
citecolor=blue,
filecolor=blue,
menucolor=blue,
urlcolor=blue,
colorlinks=true]{hyperref}

\begin{document}
\title{Planetesimal formation by sweep-up: How the bouncing barrier can be beneficial to growth}
\titlerunning{Planetesimal formation by sweep-up}
\author{F.~Windmark\inst{1,2} \and T.~Birnstiel\inst{3,4} \and C.~G\"uttler\inst{5,6} \and J.~Blum\inst{6} \and C. P.~Dullemond\inst{2} \and Th. Henning\inst{1} \authorrunning{F.~Windmark et al.}}
\institute{Max-Planck-Institut f\"ur Astronomie, K\"onigstuhl 17, D-69117 Heidelberg, Germany\\e-mail: windmark@mpia.de \and  Institut f\"ur Theoretische Astrophysik, Universit\"at Heidelberg, Albert-Ueberle-Str. 2, D-69120 Heidelberg, Germany \and Excellence Cluster Universe, Boltzmannstr. 2, D-85748 Garching, Germany  \and University Observatory, Ludwig-Maximilians University Munich, Scheinerstr. 1, D-81679, Munich, Germany \and Department of Earth and Planetary Sciences, Kobe University, 1-1 Rokkodai-cho, Nada-ku, Kobe 657-8501, Japan \and Institut f\"ur Geophysik und extraterrestrische Physik, Technische Universit\"at zu Braunschweig, Mendelssohnstr. 3, D-38106 Braunschweig, Germany}

\date{Received 18 November 2011 / Accepted 16 January 2012}

\abstract
{The formation of planetesimals is often accredited to collisional sticking of dust grains. The exact process is unknown, as collisions between larger aggregates tend to lead to fragmentation or bouncing rather than sticking. Recent laboratory experiments have however made great progress in the understanding and mapping of the complex physics involved in dust collisions.}
{We want to study the possibility of planetesimal formation using the results from the latest laboratory experiments, particularly by including the \textit{fragmentation with mass transfer} effect, which might lead to growth even at high impact velocities.}
{We present a new experimentally and physically motivated dust collision model capable of predicting the outcome of a collision between two particles of arbitrary masses and velocities. The new model includes a natural description of cratering and mass transfer, and provides a smooth transition from equal- to different-sized collisions. It is used together with a continuum dust-size evolution code which is both fast in terms of execution time and able to resolve the dust well at all sizes, allowing for all types of interactions to be studied without biases.}
{We find that for the general dust population, bouncing collisions prevent the growth above millimeter-sizes. However, if a small number of cm-sized particles are introduced, for example due to vertical mixing or radial drift, they can act as a catalyst and start to sweep up the smaller particles. At a distance of 3 AU, 100-meter-sized bodies are formed on a timescale of 1 Myr.}
{Direct growth of planetesimals might be a possibility thanks to a combination of the existence of a bouncing barrier and the fragmentation with mass transfer effect. The bouncing barrier is here even beneficial, as it prevents the growth of too many large particles that would otherwise only fragment among each other, and creates a reservoir of small particles that can be swept up by larger bodies. However, for this process to work, a few seeds of cm in size or larger have to be introduced.}

\keywords{accretion, accretion disks -- protoplanetary disks -- stars: pre-main-sequence, circumstellar matter -- planets and satellites: formation}

\maketitle

\section{Introduction}\label{sec:introduction}
One of the most popular planet formation scenarios is via core accretion, in which the formation of planets starts in the protoplanetary disk with micron-sized dust particles that collide and stick together by surface forces, forming successively larger aggregates \citep{1980PThPh..64..544M, 1996Icar..124...62P}. Traditionally, the next stage in the formation process is the gravity-aided regime where planetesimals have formed that are so massive that the gravity starts to affect the accretion and the strength of the body.

However, to reach this regime, kilometer-sized bodies are required, something which has proven difficult to produce due to a number of effects such as fragmentation and bouncing \citep{1993Icar..106..151B}, rapid inward migration \citep{Whipple:3gV1B7CM} and electrostatic repulsion \citep{2011ApJ...731...95O, 2011ApJ...731...96O}. A new planetesimal formation channel was introduced by \cite{2007Natur.448.1022J, 2011A&A...529A..62J}, in which mutual gravity plays a role already between meter-sized boulders in turbulent and locally overdense regions, resulting in a rapid formation of kilometer-sized bodies. However, even the meter regime is difficult to reach only by coagulation of dust aggregates.

The micron-sized dust particles are coupled tightly to the surrounding gas, and their relative velocities are driven primarily by Brownian motion. Since the resulting relative velocities are small, on the order of millimeters per second, the particles stick together due to van der Waals forces. However, as the particles increase in size, they become less coupled to the gas, and a number of effects increase the relative velocities between them. For centimeter-sized particles, the predicted relative velocity is already one meter per second, and meter-sized boulders collide at velocities of tens of meters per second. At these large collision energies, the particles tend to fragment rather than stick \citep{2008ARA&A..46...21B}, which effectively prevents further growth  \citep{2005A&A...434..971D, 2008A&A...480..859B, 2010A&A...513A..79B}.

In the protoplanetary disk, gas pressure supports the gas against the radial component of the stellar gravity, causing it to move at slightly sub-Keplerian velocities. Solid bodies do however not experience the supporting gas pressure, and instead drift inward. As the particles grow larger, their relative velocities compared to the gas increase, causing a significant headwind and a constant loss of angular momentum. At a distance of 1~AU, radial drift can cause meter-sized bodies to spiral inwards and get lost in the star on a timescale of a few hundred orbits \citep{1977MNRAS.180...57W, 1986Icar...67..375N}.

These two obstacles give rise to the somewhat inaccurately named meter-size barrier (which ranges from millimeters to meters depending on the disk properties), above which larger bodies have difficulties getting formed. In order to reach the gravitational regime, bodies that are roughly nine orders of magnitude more massive are needed.

The study of the dust evolution has however until recently primarily been done using simplified dust collision models in which colliding dust grains would either stick together or fragment \citep{2008A&A...480..859B, 2010A&A...513A..79B}. The simplicity of the models has been a necessity because of large uncertainties and a small parameter space covered in mass, porosity and collision velocity in the laboratory experiments and numerical simulations.

Recent years have however seen good progress in the laboratory experiments, as summarized by \cite{2008ARA&A..46...21B}. To provide a more complete and realistic collision model, \citet{2010A&A...513A..56G} reviewed a total of 19 different experiments with aggregates of varying masses, porosities and collision velocities. In these experiments, the complex outcome was classified into nine different types. \citet{2010A&A...513A..57Z} implemented this collision model in a Monte Carlo dust-size evolution code. The results showed clear differences from the previous collision models, and allowed for identification of the most important of the different collision types. They also found the important effect of dust grain bouncing at millimeter sizes that halts the grain growth even before it reaches the fragmentation barrier. With the inclusion of a vertical structure, \cite{2011A&A...534A..73Z} still found the bouncing to be prominent, but also a number of other collision effects caused by the vertical mixing.

Progress has also been made with numerical simulations of dust (silica and ice) aggregate collisions using molecular dynamics codes \citep{2009ApJ...702.1490W, 2011ApJ...737...36W} with up to 10,000 monomers corresponding to aggregate sizes of around 100 $\mu$m. Based on these simulations, \cite{2011arXiv1108.4892O} developed a collision model where growth was possible for silicates up to velocities of 7 m/s, and for ices up to 70 m/s. By implementing relative velocities in the dead zone extracted from MHD simulations, they were able to form planetesimals made of ice, but not with silicates. \cite{2010A&A...513A..58G, 2011A&A...531A.166G} also developed a dust collision code using SPH for particle sizes of cm and upwards. There is currently a discrepancy between the simulations and the laboratory experiments, where the simulations have difficulties reproducing the bouncing events and generally observe much higher fragmentation threshold velocities. We will in this paper work primarily with the (more pessimistic) laboratory data, but there is a great need to get the two fields to agree.

One possible way to grow past the fragmentation barrier is so-called fragmentation with mass transfer, which was observed by \citet{2005Icar..178..253W} and can happen in a collision between a small projectile and a large target. The projectile is fragmented during the collision and a part of it is added as a dust cone to the surface of the larger particle, provided that the mass ratio between the two particles is large enough to avoid fragmentation of the larger body. The mass transfer efficiency was studied by \citet{2010ApJ...725.1242K}, who also showed that multiple impacts over the same area still lead to growth. \cite{2009A&A...505..351T,2009MNRAS.393.1584T} have shown that growth of the target is possible even for collision velocities larger than 50 m s$^{-1}$, and \cite{2011arXiv1108.0785T} proved that the target could still gain mass even at large impact angles. These experiments have all shown that dust growth may proceed for large bodies at high velocities, and that this effect might even be able to produce planetesimals via collisional accretion. We discuss this process in more detail in Sect.~\ref{subsec:overview}.

For the study of the dust-size evolution, the Monte Carlo approach of \cite{2008ApJ...684.1291O} and \cite{2008A&A...489..931Z} has a big advantage in that it allows for the simulation of a large number of particle properties and collision outcomes. A representative particle approach is used where a few particles correspond to larger swarms of particles with the same properties. Each particle is given a set of properties, and each individual collision of the representative particles is followed. This approach uses very little computer memory, and adding extra properties cost very little in terms of execution time. If we want to study the effect of mass transfer, however, the Zsom et al. approach has some problems, as it only tracks where the most mass is in the system. It therefore has difficulties resolving wide size distributions, which is required for the type of bimodal growth that the fragmentation with mass transfer effect would result in.

Another method is the continuum approach, in which the dust population is described by a size distribution \citep{1980Icar...44..172W, 1981Icar...45..517N}. The conventional continuum approach is the Smoluchowski method, where the interactions between all particles sizes are considered and updated simultaneously. This leads to very fast codes for a one-dimensional parameter-space (i.e. mass) compared to the Monte Carlo approach. Adding further properties such as porosity and charge is however very computationally expensive in terms of memory usage and execution time if one does not include tricks like the average-porosity scheme of \cite{2009ApJ...707.1247O}. With the continuum approach, however, the dust is resolved well at all sizes, allowing for all types of dust interactions without any biases. It is also fast enough to follow the global dust evolution in the whole disk.

The aim of this paper is to create a new collision model, describing the outcome of collisions between dust aggregates of varying sizes and velocities, which is fast enough to be used with continuum codes. In this new model, we take into account the recent progress of the laboratory experiments, especially the mass transfer effect described above, and take a physical approach to transition regions from growth to erosion where the experiments are sparse. We then use this model in size evolution simulations using the local version of the code by \cite{2010A&A...513A..79B} to study its implications for the formation of the first generation of planetesimals.

The background of the new model and all the experimental work that it is based on is discussed in Sect.~\ref{sec:colmod}, and its implementation is described in Sect.~\ref{sec:implementation}. In Sect.~\ref{sec:diskmodel}, we discuss the properties of the disk in our local dust evolution simulations, as well as the implicit Smoluchowski solver that we have used. Finally, in Sects.~\ref{sec:results} and \ref{sec:discussion}, we discuss the results from the new model and show how the existence of a bouncing barrier may even be beneficial to the growth of planetesimals.

\section{Motivation for the new collision model}\label{sec:colmod}
Models to describe the growth of dust aggregates can generally be divided into two parts: A \textit{collision model} describes the result of a collision between two dust particles of arbitrary properties (i.e. mass, porosity) and velocities. A \textit{dust evolution model} uses the collision model to describe the evolution of the particle properties of an entire population of dust particles as they collide and interact with each other. In this section, we describe the latest laboratory experiments and our effort to produce a collision model that can take these results into account while still streamlining it to work well with continuum dust evolution codes. That means that the collision model cannot be as complex as the one developed by \citet{2010A&A...513A..56G}, but needs to focus on the most important collision types and aggregate properties. Still, we were able to include results which were not established or even known when the model of G\"uttler et al. was developed.

\subsection{Overview of recent experiments and simulations}
\label{subsec:overview}

Numerous laboratory experiments have been performed to probe the collision parameter space of silicate dust grains, as summarized by \cite{2008ARA&A..46...21B}. This is a daunting task, as planet formation spans more than 40 orders of magnitude in mass and 6 orders of magnitude in collisional velocity and collisional outcomes are affected by for example porosity, composition, structure and impact angle. The classical growth mechanism of dust grains is the hit-and-stick mechanism, which has been well studied both in laboratory experiments \citep[BW00]{2000Icar..143..138B} and in numerical simulations \citep{1997ApJ...480..647D, 2009ApJ...702.1490W}. Sticking collisions are also possible via plastic deformation at the contact zone \citep[WGB12]{2011arXiv1105.3909W} and geometrically by penetration \citep[LTB08]{2008ApJ...675..764L}.

Previous collision models have due to limited data with few exceptions only included sticking, cratering and fragmentation with simplistic thresholds \citep{1986Icar...67..375N, 1997Icar..127..290W, 2005A&A...434..971D, 2005ApJ...625..414T, 2008A&A...480..859B}. In order to study the effect of the progress in laboratory experiments, \cite{2010A&A...513A..56G} and \cite{2010A&A...513A..57Z} presented a collision model containing nine different collisional outcomes and used this in the Monte Carlo dust evolution code developed by \cite{2008A&A...489..931Z}. Their model contained three additional types of sticking collisions besides the normal hit-and-stick, and they also identified two growth-neutral bouncing effects and three different fragmentation effects in which the largest particle is eroded. It was found that several of the new collision types played a role for the dust-size evolution, which proved the necessity for a more complex dust collision approach than what had been previously used. Before even reaching the fragmentation barrier, at which fragmentation events between similar-sized particles prevent further growth, they found the existence of the so-called bouncing barrier. Bouncing collisions between smaller particles of intermediate velocities proved to be an efficient barrier for growth already at millimeter-sizes. It should be clarified that bouncing is, in principle, not bad for growth. What makes the bouncing barrier a problem is the lack of sticking over such a large range of masses and velocities that there is no way for the particles to grow further.

Bouncing between dust aggregates is at the moment a hotly discussed topic. It has been reported from a large number of laboratory experiments of different setups and material properties \citep{1993Icar..106..151B, 2007epsc.conf..840H, 2008ApJ...675..764L, 2009PhRvL.103u5502K, 2010A&A...513A..56G, 2011arXiv1105.3909W}, but molecular dynamics simulations show significantly less or no bouncing \citep{2007ApJ...661..320W, 2008ApJ...677.1296W, 2009ApJ...702.1490W, 2009A&A...507.1023P}. These rebounding events happen in collisions where the impact energy is so high that not all can be dissipated by restructuring of the aggregates. \cite{2011ApJ...737...36W} argue that this would happen only for very compact aggregates where the coordination number is high, which is in contradiction to what is seen in the laboratory. We will in this work base our model on the laboratory experiments, but this is a very important matter for the dust growth and will need to be investigated further.

\begin{figure}[t]
\centering
\resizebox{0.9\hsize}{!}{\includegraphics{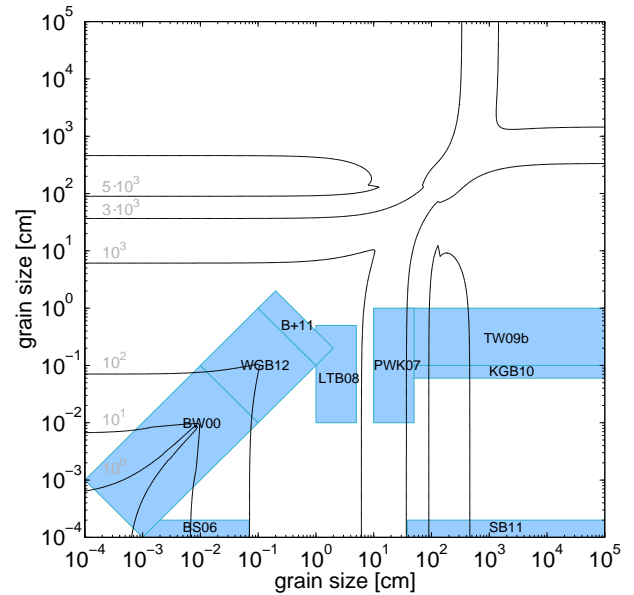}}
\caption{The size-size parameter space of the dust collision experiments (blue boxes) laying basis for the new collision model. All marked experiments are discussed in more detail in Sect.~\ref{subsec:overview}. The contours and gray labels mark the collision velocities in cm~s$^{-1}$ expected from the disk model described in Sect.~\ref{subsec:diskmodel}, and do not always coincide with the velocities studied in the experiments.}
\label{fig:experimentmap}
\end{figure}

In our new model, we implement the most important collision types identified by \cite{2010A&A...513A..57Z}, and also take into account the results from a number of recent experimental studies. Many new experiments have been performed that have further increased the understanding of the collision physics of dust aggregates. In Fig.~\ref{fig:experimentmap}, we plot the parameter space of a selection of important laboratory experiments laying basis for the new collision model.

Provided that the mass-ratio between the two involved particles is large enough (from now on called the projectile and the target for the smallest and largest particle, respectively), the projectile can fragment and parts of it stick by van der Waal forces to the surface of the target. This was studied by \cite{2005Icar..178..253W} and \citet[TW09b]{2009MNRAS.393.1584T} for millimeter to centimeter-sized projectiles shot on a mounted decimeter-sized dust target at velocities up to 56.5 m s$^{-1}$. It was found that the accretion efficiency even increased with velocity, and could be as high as 50\% of the mass of the fragmented projectile, where \citet{2010A&A...513A..56G} only assumed a constant 2\%. The effect was also observed by \citet[PWK07]{2007Icar..191..779P} in drop tower experiments where also the target was free-floating without a supported back. The mass transfer efficiency at slightly lower velocities (1.5 - 6 m s$^{-1}$) and with millimeter-sized projectiles was studied in more detail by \citet[KGB10]{2010ApJ...725.1242K}, who confirmed the velocity-positive trend. \cite{2009A&A...505..351T} and Kothe et al. also studied multiple impacts over the same area, and could conclude that growth was possible even then, without the newly accreted material being eroded. It was also found that growth was possible even at very steep impact angles. \citet[B+11]{2011ApJ...736...34B} performed experiments between cm-sized particles at even lower velocities (8 mm s$^{-1}$ to 2 m s$^{-1}$) and found mass transfer events already at 20 cm s$^{-1}$, right at the onset of fragmentation.

This mode of growth, where small projectiles are impacting large targets, is related to the work of \cite{2003EP&S...55..263S, 2005Icar..176..220S}. They did numerical studies on whether small fragments formed in an erosive collision could be reaccreted back onto the target due to gas drag. The conclusion was that if the fragments were $\mu$m-sized, and the target large enough, the gas flow would actually lead the fragments around the target and thereby prevent reaccretion. For the mass transfer effect, it is important to verify whether this effect could prevent 100-1000 $\mu$m-sized projectiles from impacting on the target in the first place. The importance of this effect can be estimated with a simple comparison of timescales using reasonable parameters from the disk model (discussed in more detail in Sect.~\ref{sec:diskmodel}). The stopping time of a small particle is given by
\begin{equation}
	\tau_{\rm s} = \frac{ \xi a_{\rm p}}{ \rho_g \cdot \bar{u}} \sim 2500~~~{\rm s}
\end{equation}
where $\xi \sim 1$ g cm$^{-3}$ is the solid density of the projectile, $a_{\rm p} \sim 100$ $\mu$m is its radius, $\rho_g \sim 10^{-10}$ g cm$^{-3}$ the expected midplane gas density at 3 AU and $\bar{u} \sim 4 \cdot 10^4$ cm s$^{-1}$ the mean thermal velocity of the gas. The time it would take for the projectile to pass the target is given by
\begin{equation}
	\tau_{\rm pass} = \frac{a_{\rm t}}{\Delta v} \sim 0.2~~~{\rm s}
\end{equation}
where $a_{\rm t} \sim 100$ cm is a typical target size and $\Delta v \sim 5000$ cm s$^{-1}$ the relative velocity between the particles. Since $\tau_{\rm s} >> \tau_{\rm pass}$, it would take too long for the projectile to adjust to the gas flow around the target, and the two particles would collide. If the projectile was instead only 1 $\mu$m in size, the timescales would not differ so much, and the gas flow might play a role.

Another recent experimental progress is the refinement of the threshold velocity for destructive fragmentation, where the target is completely disrupted. \cite{2011ApJ...736...34B} performed experiments to determine the onset of global fragmentation of the particles, and found that cm-sized particles fragmented already at 20 cm s$^{-1}$, much below the 1 m s$^{-1}$ threshold found for mm-sized particles by \cite{1993Icar..106..151B}. This points towards a material strength that decreases with mass, as predicted for rocky materials among others by \cite{1999Icar..142....5B}. This can be explained by a larger probability for faults and cracks in the material the larger the particle, and it is along these cracks that global breaking and fragmentation takes place. No experiments have as of yet been performed to study the fragmentation threshold between different-sized dust aggregates, but one can generally assume that the velocities needed would be higher with an increasing size-ratio, as is seen both in experiments and simulations of collisions between rocky materials \citep{2011arXiv1109.4588S, 2012ApJ...745...79L}.

In order to provide more data in the transition region between sticking and bouncing, \citet{2011arXiv1105.3909W} studied collisions between particles of $0.5 - 2$ mm in size and at velocities of $0.1-100$ cm s$^{-1}$. In these experiments, sticking collisions were found (in coexistence with bouncing events) for higher velocities than was previously expected \citep{2000Icar..143..138B, 2010A&A...513A..56G}, and enough data now exists to define a transition regime between only sticking and bouncing. Similar experiments with smaller particles roughly 100 $\mu$m in size were performed by Kothe et al. (unpublished), and are consistent with the threshold of Weidling et al.

\citet[SB11]{2011ApJ...734..108S} have also performed erosion experiments between $\mu$m-sized monomer projectiles and mounted high-porosity aggregates for velocities up to 60 m s$^{-1}$, in order to determine the erosion efficiency as a function of the collision velocity and the surface structure. They found the initial stages of the monomer bombardment to be very efficient even at low velocities, but after the most loosely bound monomer chains had been knocked off and the surface had compacted, the erosion was greatly decreased.

\subsection{Individual treatment of collisions}

In the collision model of \cite{2010A&A...513A..56G}, a binary approach was used for the particle mass ratios and porosities. Below a certain set critical mass ratio, $r_{\rm c} = m_{\rm t} / m_{\rm p}$ = 10, 100, 1000, the collision was treated as being between equal-sized particles, leading for example to global fragmentation if two large particles collided at high velocities. If the mass ratio was above the critical ratio, the particles were treated as different-sized, and a high-velocity collision would instead lead only to cratering. The same approach was taken for the porosity. Below a critical porosity $\phi_{\rm c} = 0.4$, a particle was considered porous for the purpose of determining the collision outcome, and above it, the particle was considered compact. Combining these two binary properties gave eight different collision scenarios, where the collision outcome was determined by the particles masses, porosities and relative velocity.

In the new model, we instead use the current laboratory data to do an interpolation between the two extreme mass-ratios. This provides a continuous transition from equal-sized to different-sized collisions, and allows us to distinguish between collisions of different sizes at intermediate mass-ratios, and provides a natural and smooth transition between the two extremes. We can therefore obtain what velocity is needed to cause global fragmentation for a specific mass-ratio, which gives us a more precise tool to study the effects of where  global fragmentation turns into local cratering.

It is however necessary for us to make a simplification regarding the porosity of the dust grains. Adding additional properties to the dust grains is very computationally expensive for continuum codes like the Smoluchowski solver that we use for the dust-size evolution, compared to Monte Carlo codes. In the Monte Carlo approach, each timestep only consists of one collision between a representative particle and a swarm of identical particles. After the collision, the properties (i.e. mass, porosity, charge) of the representative particle is updated, and a new timestep is initiated. This means that for a simulation with $n$ representative particles, each new property only adds an additional time $O(n)$ to the execution time.

In the Smoluchowski method, one has to numerically solve a number of differential equations to update the number density of all mass bins. For each grain size, $n^2$ interaction terms need to be considered, where $n$ is the number of mass bins. This is because a mass bin can collide with all bins including itself, but fragmentation can cause mass to be put into it also from a collision between two other bins. If an additional property such as porosity is included, $m = n$ porosity bins would need to be included. For each $n \cdot m$ bin, $(n \cdot m)^2$ interactions would now need to be considered, and the code would be slower by a factor $O(m^3)$. In order to include porosity in the Smoluchowski solver, we would therefore require some analytical trick like an average porosity for each mass bin described in \cite{2009ApJ...707.1247O}. This is however outside the scope of this paper, and we instead assume that all particles are compact at all times. This is likely a good approximation for larger particles outside the hit-and-stick region, as bouncing collisions quickly lead to compaction of the particles. This finally gives us one single collision scenario, where we can for a collision between any two given particles determine the outcome based on their masses and relative velocity.

\section{Implementation of the model}\label{sec:implementation}
\begin{figure*}[t]
\centering
\includegraphics[width=17cm]{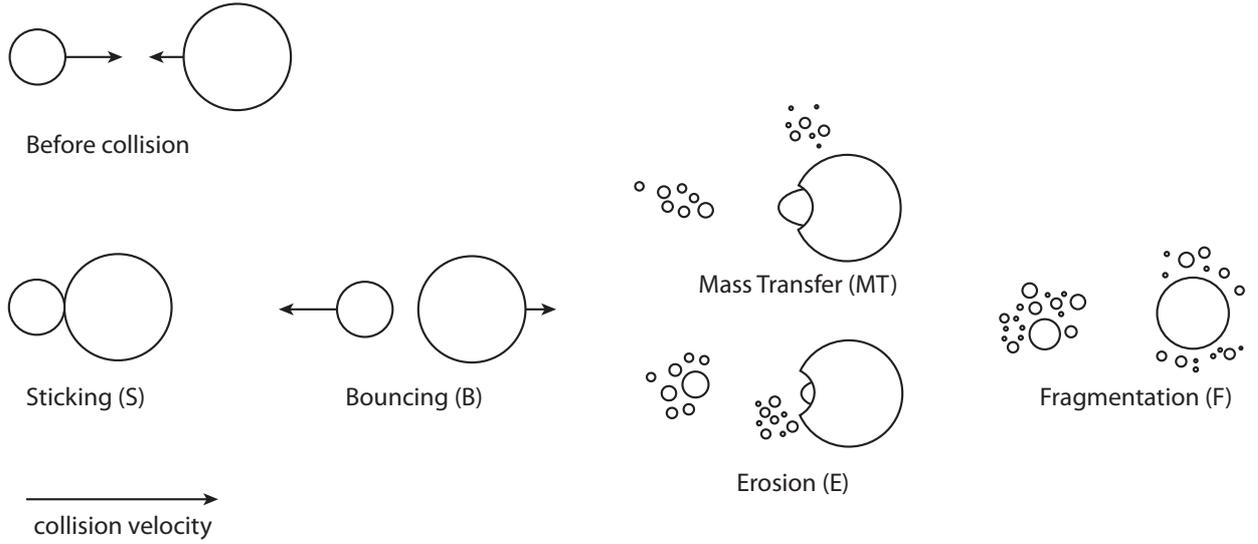}
\caption{Sketch of the five possible outcomes described in Sect.~\ref{sec:implementation} sorted in rough order of collision velocity. Mass transfer and erosion act simultaneously in a collision, and we define a mass transfer collision as leading to net growth for the target, and an erosive collision leading to net mass loss. This outcome is extremely dependent on the mass-ratio between the particles, thus adding a second, vertical dimension to the sketch.}
\label{fig:outcomesketch}
\end{figure*}

We discuss in this section the details on how the new collision model has been created and implemented into the code. We choose to include only the collision types that proved to be most important in the simulations of \citet{2010A&A...513A..57Z}. The collision types considered here are sticking, bouncing and the transition between them, mass transfer combined with erosion, and destructive fragmentation. These are shown schematically in Fig.~\ref{fig:outcomesketch}, and are discussed in detail in Sects.~\ref{subsec:general}-\ref{subsec:mtander}. In Table~\ref{tab:collparams}, we give a summary of all the symbols that are used in this section.

\subsection{Sticking and bouncing thresholds}
\label{subsec:general}

We consider two dust grains colliding with a relative velocity $\Delta v$. The projectile has a mass $m_p$ and the target a mass $m_t \geq m_p$. \cite{2011arXiv1105.3909W} found mass-dependent sticking and bouncing threshold velocities that can be written as
\begin{equation}
	\label{eq:stick}
	\Delta v_{\rm stick} = \left( \frac{ m_{\rm p} }{ m_{\rm s}} \right)^{-5/18}~~~ [{\rm cm~s^{-1} }]
\end{equation}
and
\begin{equation}
	\label{eq:bounce}
	\Delta v_{\rm bounce} = \left( \frac{ m_{\rm p} }{ m_{\rm b} } \right)^{-5/18}~~~ [{\rm cm~s^{-1} }]
\end{equation}
where $m_{\rm s} = 3.0 \cdot 10^{-12}$ g and $m_{\rm b} = 3.3 \cdot 10^{-3}$ g are two normalizing constants calibrated by laboratory experiments, and the $\Delta v \propto m^{-5/18}$ proportionality is consistent with the theoretical models of \cite{Thornton:1998ve}. The above two thresholds mean that collisions with $\Delta v < \Delta v_{\rm stick}$ result in 100\% sticking, and $\Delta v > \Delta v_{\rm bounce}$ result in 100\% bouncing (provided that neither of the particles involved are fragmented). Between these two thresholds, we have a region where both outcomes are possible, described in more detail in Sect.~\ref{subsec:outcomes}.

\begin{table}[!b]
\caption{Symbols used in the collision model}
\begin{tabular}{ l l}
\hline
\hline
Symbol & Meaning \\
\hline
$a_{\rm p/t}$			& radius of the projectile/target \\
$m_{\rm p/t}$			& mass of the projectile/target \\
$\Delta v$				& relative velocity between the particles \\
$\Delta v_{\rm stick}$	& sticking threshold velocity \\
$m_{\rm s}$			& sticking threshold normalizing constant \\
$\Delta v_{\rm bounce}$	& bouncing threshold velocity \\
$m_{\rm b}$			& bouncing threshold normalizing constant \\
$v_{\rm p/t}$			& center-of-mass velocity of the projectile/target \\
$\mu$				& relative mass of the largest remnant \\
$m_{\rm rem}$			& mass of the largest remnant \\
$m_{\rm mt}$			& mass transferred from the projectile to the target \\
$\epsilon_{\rm ac}$		& accretion efficiency during mass transfer \\
$m_{\rm er}$			& mass eroded from the target due to cratering \\
$\Delta m_{\rm t}$		& net mass change from mass transfer and erosion \\
$\epsilon_{\rm net}$		& net accretion efficiency from mass transfer and erosion \\
$v_{\mu}$				& velocity needed to fragment with largest remnant $\mu$ \\
$m_{\mu}$			& fragmentation threshold normalizing constant \\
$m_0$				& mass of a monomer ($= 3.5\cdot 10^{-12}$ g) \\
$m_{\rm frag}$			& total mass of the fragments \\
\hline
\end{tabular}
\label{tab:collparams}
\end{table}

\subsection{An energy division scheme for fragmentation}

From the fragmentation with mass transfer experiments described in the previous section, we make the assumption that mass transfer at a varying efficiency occurs in all cases where the projectile fragments. If also the target fragments, the mass transfer is negligible compared to the huge mass loss, and we can safely ignore it. We therefore need to determine for each collision whether one, both or neither of the particles fragment.

The majority of the dust collision experiments have however been performed between either equal-sized or very different-sized particles. In order to interpolate between these two extremes, we need to look at the collision energy of the event, and determine how this energy is distributed between the two particles. Not only the collision energy of an event matters when determining the degree of fragmentation, but also the mass-ratio between the two particles. In two collisions with equal collision energy but with different mass-ratios, we expect the higher mass-ratio collision to be less efficient in completely disrupting the target, as the energy will be more locally distributed around the contact point. In order to take this into account, we choose to look at the particles in the center-of-mass frame. In this frame, the massive particle is moving more slowly than the small, and during the collision moment, the kinetic energy of the particles will be reduced to zero. Physically, this corresponds to a fully plastic collision where all the energy is consumed by deformation and fragmentation.

We do in this approach assume that the kinetic energy of each particle in the center-of-mass frame will be used to try to fragment itself. The velocities for the two particles in the center-of-mass frame are given by
\begin{align}
	v_{\rm p} &= \frac{\Delta v}{ 1 + m_{\rm p} / m_{\rm t }} \\
	v_{\rm t} &= \frac{\Delta v}{ 1 + m_{\rm t} / m_{\rm p }}
\end{align}
All velocities in the center-of-mass frame will from now on be denoted as $v$ and will then mean either $v_p$ or $v_t$. The above equations mean that the largest particle will have the lowest velocity in the center-of-mass frame. In the case with an extreme mass ratio, $m_{\rm p} / m_{\rm t} \rightarrow 0$, the center-of-mass velocity of the projectile and target is given by $v_{\rm p} = \Delta v$ and $v_{\rm t} = 0$, respectively.

During a fragmenting collision, the relative size of the largest remnant can be described by
\begin{equation}
	\mu_{\rm p/t} = \frac{ m_{\rm rem} }{ m_{\rm p/t} }
\end{equation}
where $m_{\rm rem}$ is the mass of the largest remnant and $m_{\rm p/t}$ is the original particle mass. Depending on their sizes and material strengths, the two original particles can be fragmented to different degrees. In this model, each collision partner is treated individually with a $\mu_{\rm t}$ and $\mu_{\rm p}$ for the remnant of the target and the projectile, respectively. We define the center-of-mass velocity required for the largest remnant to have a relative mass $\mu$ as $v_\mu$.

\cite{1993Icar..106..151B} and \cite{Lammel:mLrgVib9} studied the threshold velocities needed for two mm-sized particles to fragment with largest remnants of relative masses $\mu = 1.0$ and $\mu = 0.5$, where the former corresponds to the onset of fragmentation and the latter corresponds to a largest remnant equal to half of the original particle. \cite{2011ApJ...736...34B} studied the threshold velocities for cm-sized particles. Interpolating between the results for the two sizes, the center-of-mass frame threshold velocity can be written as
\begin{equation}
	\label{eq:frageq}
	v_{\mu} = (m / m_\mu)^{-\gamma}~~~ [{\rm cm~s^{-1} }]
\end{equation}
where $m_\mu$ is a normalizing constant calibrated by the laboratory experiments and $\gamma = 0.16$. The fragmentation threshold velocity is given by $v_{1.0}$, with $m_{1.0} = 3.67 \cdot 10^{7}$ g. The velocity required for the largest fragment to have half the size of the original particle is $v_{0.5}$, with $m_{0.5} = 9.49 \cdot 10^{11}$ g. The relative mass of the largest fragment is fitted to a power law dependent on velocity and mass:
\begin{equation}
	\label{eq:mueq}
	\mu(m,v) = C \cdot \left( \frac{m}{1~{\rm g}} \right)^\alpha \cdot \left( \frac{v}{1~{\rm cm~s}^{-1}}\right)^\beta
\end{equation}
The above equation is valid for all velocities $v > v_{1.0}$. By fitting the $\mu(m,v)$ plane to the two parallel threshold velocities given by Eq.~(\ref{eq:frageq}), we get
\begin{align}
	\alpha &= \log(2) / \log(m_{1.0}/m_{0.5}) = -0.068 \\
	\beta &= \alpha / \gamma = -0.43\\
	C &= m_{1.0}^{-\alpha} =  3.27~~~[{\rm g}^{-\alpha}]
\end{align}
This means that at a larger collision velocity, the particle will fragment more heavily and the size of the largest fragment will decrease. More mass is therefore put into the lower part of the mass spectrum.

We can from Eq.~\ref{eq:mueq} determine the largest fragment for each of the particles in the collision, and also use it to identify fragmenting collisions. If $\mu_p < 1$ and $\mu_t \geq 1$, only the projectile fragments and mass transfer occurs. If both $\mu_p < 1$  and $\mu_t < 1$, both particles fragment globally. Since the center-of-mass velocity $v$ is inversely proportional to the mass of the particle, we will never have a case where only the target fragments and the projectile is left intact, even if $v_\mu$ decreases with mass.

\subsection{A new mass transfer and cratering model}
\label{subsec:mtander}

\begin{figure}[t]
\centering
\resizebox{0.6\hsize}{!}{\includegraphics{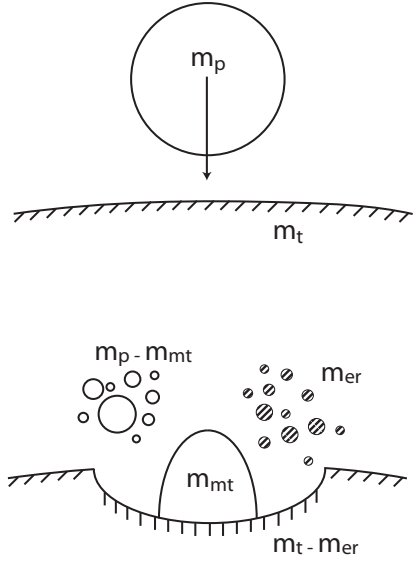} }
\caption{ Sketch of the combined cratering and mass transfer process, which occurs during a high-enough velocity collision between a projectile with mass $m_{\rm p}$ and a target with mass $m_{\rm t}$. During the collision, an amount $m_{\rm mt}$ is added from the projectile to the target, and the rest of the projectile is put into small fragments with a total mass $m_{\rm p,frag}$. An amount $m_{\rm er}$ is simultaneously eroded from the target, and the final mass of the target is given by $m_{\rm t}' = m_{\rm t} + m_{\rm mt}  - m_{\rm er}$.}
\label{fig:cratersketch}
\end{figure}

We use a new realistic approach to distinguish between collisions where the target experiences net mass gain due to mass transfer, and where it experiences net mass loss due to cratering. During each collision, we assume that there is simultaneously:
\begin{itemize}
\item mass added to the target from the projectile via mass transfer
\item mass eroded from the target due to cratering
\end{itemize}
We also assume that these two effects act independent of each other. This is illustrated in Fig.~\ref{fig:cratersketch}, and can be seen in the high-velocity experiments by \cite{2009MNRAS.393.1584T}. The mass change of the largest particle can be described by:
\begin{equation}
	\label{eq:masschange}
	\Delta m_{\rm t} = m_{\rm mt} - m_{\rm er}
\end{equation}
where $m_{\rm mt} = \epsilon_{\rm ac} \cdot m_{\rm p}$ is the mass added due to mass transfer with the accretion efficiency $0 \leq \epsilon_{\rm ac} \leq 1$ and $m_{\rm er}$ is the mass lost due to cratering. An increasing velocity not only leads to increased mass transfer, but also increased cratering. This makes it possible to naturally determine where growth transitions into erosion.

The mass transfer efficiency is obtained from \cite{2011ApJ...736...34B}, and depends on both the particle porosity and velocity. Since we are unable to track the porosity of the particles, we assume a constant porosity difference of $\Delta \phi = 0.1$ between the two dust aggregates, where the target always is the more compact one. This is likely a reasonable approximation for larger particles which have left the hit-and-stick phase and have had time to compact during bouncing collisions, which is the region where mass transfer can be expected. In our prescription, we also include a fragmentation threshold velocity dependence, so that the efficiency is always the same for the same degree of projectile fragmentation. This results in
\begin{equation}
	\label{eq:masstransfer}
	\epsilon_{\rm ac}   = -6.8 \cdot 10^{-3} + 2.8 \cdot 10^{-4} \cdot \frac{v_{\rm 1.0,beitz}}{v_{1.0}} \cdot \frac{\Delta v}{1~{\rm cm~s^{-1}}}
\end{equation}
where $v_{\rm 1.0,beitz} = 13$ cm s$^{-1}$ is the onset of the fragmentation for the 4.1 g particles used by \cite{2011ApJ...736...34B}, and $v_{1.0}$ is the fragmentation threshold calculated for the mass of the projectile, both calculated using Eq.~\ref{eq:frageq}. We here take a maximum mass transfer efficiency of $\epsilon_{\rm ac} = 0.5$ as indicated by \cite{2005Icar..178..253W}. Due to the process of fragmentation and mass transfer considered here, a higher value would not be reasonable as it would point towards complete sticking, which was never observed at these velocities.

If the collision energy is not high enough to fragment the particle globally, some of the energy is still used to break up local bonds between monomers around the contact point, resulting in cratering. The cratering efficiency has however only been studied in a couple of laboratory experiments. \cite{2011ApJ...734..108S} found for monomer projectiles an erosion efficiency given by
\begin{equation}
	\frac{m_{\rm er}}{m_{\rm p}} = 1.55\cdot 10^{-4} \cdot \frac{\Delta v}{1~{\rm cm~s^{-1}}} - 0.4
\end{equation}
where $m_{\rm er}$ is the amount of eroded mass and $m_{\rm p} = m_0$ is the projectile mass, and $m_0 = 3.5 \cdot 10^{-12}$ g is the monomer mass. \cite{2007Icar..191..779P} studied the erosion of porous targets both with solid and porous projectiles, and found results that varied highly with the porosity of the projectile and target. Their results are therefore highly uncertain, but roughly agree with an erosion efficiency of
\begin{equation}
	\frac{m_{\rm er}}{m_{\rm p}} = \frac{3}{400} \cdot  \frac{\Delta v}{1~{\rm cm~s^{-1}}}
\end{equation}
\begin{figure}[t]
\centering
\resizebox{1\hsize}{!}{\includegraphics{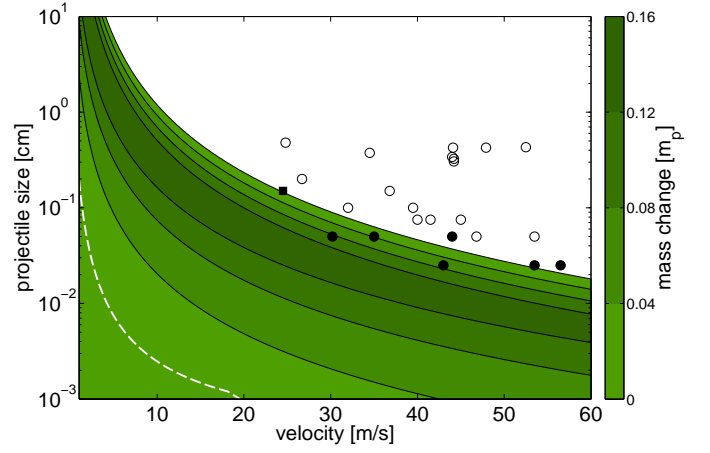} }
\caption{The threshold between growth and erosion from the model compared to the mass-transfer experiments performed by \cite{2009MNRAS.393.1584T}. Filled circles show experiments where the target gained mass, and open circles show where it lost mass. The white dotted line shows the threshold from the highly uncertain erosion prescription from cratering experiments (collisions above the line result in erosion, and collisions below result in growth). The contours are in intervals of 4\% net accretion efficiency and mark the region with net growth in the final prescription calibrated using the Teiser \& Wurm data.}
\label{fig:craterbalance}
\end{figure}
It should however be noted that for the more compact dust aggregates that are expected after the compression by bouncing phase, the erosion efficiency should be greatly decreased, as is generally seen by \cite{2009MNRAS.393.1584T}. To interpolate between the two experiments where the degree of erosion has been measured, we assume a mass power-law dependency which can be written as
\begin{equation}
	\label{eq:erosion}
	\frac{m_{\rm er}}{m_p} = a \cdot \left( \frac{m_{\rm p}}{m_0} \right)^k \cdot \frac{\Delta v}{1~{\rm cm~s^{-1}}} + b
\end{equation}
where $a$, $b$ and $k$ are fitting parameters. The above two erosion experiments give the efficiency of two different physical effects. In monomer impacts, the projectile hits single surface monomers and sometimes manages to break the bonds between a couple of them. For larger projectiles, restructuring of the target absorbs a lot of the collision energy, and a crater is formed both because of surface compaction and breaking of monomer bonds. Direct comparisons and interpolations between the efficiencies of the two effects can not be done without huge uncertainties. A direct interpolation between the two effects yields $a = 1.55 \cdot 10^{-4}$, $b = -0.4$ and $k = 0.14$, but we will below present another way to obtain a reasonable erosion prescription.

As was discussed in the previous section, during a collision, erosion and mass transfer usually occurs simultaneously, and the net mass change of the target is given by Eq.~\ref{eq:masschange}. For $\Delta m_{\rm t} > 0$, the target experiences net growth, and for $\Delta m_{\rm t} < 0$, the target experiences net erosion. With this prescription, the transition region is extremely sensitive to the efficiency of the erosion. 

In Fig.~\ref{fig:craterbalance}, we plot the results from the mass transfer experiments performed by \cite{2009MNRAS.393.1584T}. We compare this to the threshold between growth and erosion ($\Delta m_{\rm t} = 0$) obtained from the mass transfer prescription of Eq.~\ref{eq:masstransfer} and the erosion prescription of Eq.~\ref{eq:erosion}. The resulting threshold using the experiment erosion interpolation is given by the white dashed line, and is very pessimistic compared to the mass-transfer experiments.

Since the experimental erosion prescription is obtained from a very different parameter space than what we are interested in, it is highly uncertain, and much more so than the mass transfer experiments discussed below. We therefore choose to calibrate the three parameters $a$, $b$ and $k$ of Eq.~\ref{eq:erosion} using the experimentally obtained threshold between growth of erosion of \cite{2009MNRAS.393.1584T}. This results in 
\begin{align}
	a &= 1.1 \cdot 10^{-5} \nonumber \\ 
	b &= -0.4 \\
	k &= 0.15 \nonumber
\end{align}
Comparing the net growth efficiency from this fit marked by the contours in Fig.~\ref{fig:craterbalance} to the mass transfer experiments of \cite{2010ApJ...725.1242K} (with 1 mm projectiles at velocities of 1-6 m s$^{-1}$) and \cite{2005Icar..178..253W} (with 1-10 mm projectiles up to 25 m s$^{-1}$) results in a rough agreement, even though our model ends up slightly pessimistic compared to their results, with net efficiencies roughly half of theirs. Regardless of this discrepancy, we take this conservative estimate of the experiments and use it for for our model.

\subsection{Fragmentation distribution}
\label{subsec:fragments}

During cratering, mass transfer and destructive fragmentation events, the mass of each fragmented particle is divided into two parts; the power-law distribution and the largest fragment. The fragment power-law was determined experimentally by \cite{1993Icar..106..151B} and also used in the model of \cite{2010A&A...513A..56G}, and is written
\begin{equation}
	\label{eq:fragdist}
	n(m) {\rm d}m  \propto m^{-\kappa}~{\rm d}m
\end{equation}
where $n(m) {\rm d}m$ is the number density of fragments in the mass interval $[m, m + {\rm d}m]$, and $\kappa = 9/8$.

If the mass of the largest remnant is given by $\mu \cdot m$, where $\mu$ is the relative size of the largest remnant described by Eq.~\ref{eq:mueq}, the total mass that is put into the power-law distribution is equal to $(1-\mu) \cdot m$. We give the upper limit of the fragmentation distribution by min$[(1-\mu),\mu] \cdot m$. This means that as long as $\mu < 0.5$, we have a single distribution up to the largest remnant. For $\mu > 0.5$, on the other hand, more than half of the mass is put into the largest remnant, which is then detached from the power-law distribution.

\begin{figure}[t]
\resizebox{1.0\hsize}{!}{\includegraphics{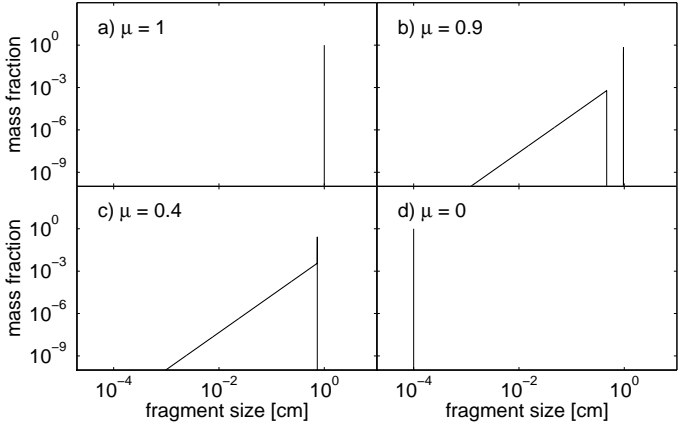} }
\caption{ The fragment mass distribution for a 1 cm-sized particle after destructive fragmentation events of varying degrees. The largest remnant is equal to in a) $\mu = 1$, b) $\mu = 0.9$, c) $\mu = 0.5$, d) $\mu = 0$ in units of the original particle mass.}
\label{fig:fragdist}
\end{figure}

This fragmentation recipe is similar to the four-populations model of \cite{2011A&A...531A.166G}, with the difference that we treat the fragmentation of both particles individually. It is able to describe all different degrees of fragmentation, and in Fig.~\ref{fig:fragdist}, we illustrate the fragment distribution for four different values of $\mu$. In a), we are at the onset of the fragmentation, and all of the mass is put back into the remnant, leading to no erosion. In b), more than half of the mass is put into the remnant, which is therefore detached from the size distrubution, and in c), the erosion is so strong that that remnant has attached to the power-law distribution. Finally, in d), the particle is completely pulverized, and all of the mass is put into monomers.

\subsection{Implementation into the code}
\label{subsec:outcomes}

In this section, we summarize the conditions and outcome for each individual collision type as they have been implemented into the code. The different types are, in order, sticking, transition from sticking to bouncing, bouncing, mass transfer combined with erosion, and destructive fragmentation, and they are all shown schematically in Fig.~\ref{fig:outcomesketch}. The conditions for sticking and bouncing are given in Eqs.~\ref{eq:stick} and \ref{eq:bounce}, and we use Eq.~\ref{eq:mueq} to determine which, if any, of the particles get fragmented during a collision, resulting in fragmentation with mass transfer or destructive fragmentation.
\\
\\
{\bf  Sticking:} ($\Delta v < \Delta v_{\rm stick}$)
\\
The two particles stick together and form a bigger particle with size $m_{\rm big} = m_{\rm t} + m_{\rm p}$.
\\
\\
{\bf Sticking/bouncing transition:} ($\Delta v < \Delta v_{\rm bounce}$)
\\
Transition from 100\% sticking to 100\% bouncing. We assume a logarithmic probability distribution between $\Delta v_{\rm stick} < \Delta v < \Delta v_{\rm bounce}$ given by
\begin{equation}
	p_{\rm c} = 1 - k_1 \cdot \log_{10}(\Delta v) - k_2
\end{equation}
where $p_{\rm c}$ is the coagulation probability. We know that at the sticking threshold (Eq.~\ref{eq:stick}), the sticking probability is $p_{\rm c} = 1$, and at the bouncing threshold (Eq.~\ref{eq:bounce}), the coagulation probability is $p_{\rm c} = 0$. The constants are then
\begin{align}
	k_1 &= \frac{18/5}{\log_{10}(m_{\rm b} / m_{\rm s} )} = 0.40\\
	k_2 &= \frac{ \log_{10}( m_{\rm p} / m_{\rm s}) }{ \log_{10}(m_{\rm b} / m_{\rm s}) }
\end{align}
{\bf Bouncing:} ($\Delta v > \Delta v_{\rm bounce}$), ($\mu_p > 1$) and ($\mu_t > 1$), ($\Delta v < v_{\rm er}$)
\\
If the collision energy is too high to result in a sticking collision but too low to fragment or erode any of the particles, the collision results in a growth-neutral bouncing event. The two masses involved in the collision are left unchanged. This type of collision results in compaction of both particles, but we ignore any porosity changes in this model.
\\
\\
{\bf Mass transfer/Erosion:} ($\mu_p < 1$) and ($\mu_t > 1$) or ($m_{\rm er} > 0$)
\\
If the collision velocity is high enough, erosion of the target will occur (Eq.~\ref{eq:erosion}). Simultaneously, if only the projectile fragments, we have a fragmentation with mass transfer event (Eq.~\ref{eq:masstransfer}). The resulting mass change of the target is given by Eq.~\ref{eq:masschange}.

The fragmented mass from the projectile is divided into two parts, a power-law and the largest remnant, with a total mass of $m = (1-\epsilon_{\rm ac})m_{\rm p}$. The power-law distribution has a total mass of $m_{\rm frag} = (1-\epsilon_{\rm ac})(1-\mu_{\rm p})m_{\rm p}$ and the largest fragment a mass $m_{\rm rem} = (1-\epsilon_{\rm ac})\mu_{\rm p}m_{\rm p}$. The fragments excavated from the target due to the cratering are also distributed after the power-law described in Sect.~\ref{subsec:fragments}, with an upper limit equal to $m_{\rm er}$.
\\
\\
{\bf Fragmentation:} ($\mu_p < 1$) and ($\mu_t < 1$)
\\
Finally, if the collision velocity is high enough and the mass-ratio is not too large, we get a destructive fragmentation event where both particles are fragmented. We treat the fragmentation of each particle individually, and get two separate fragment distributions, one for the projectile and one for the target. Each distribution is divided into two parts; the fragmentation power-law distribution with a total mass of $m_{\rm frag} = (1-\mu_{\rm p/t})m_{\rm p/t}$ and the largest fragmentation remnant with a mass of $m_{\rm rem} = \mu_{\rm p/t} m_{\rm p/t}$.

\section{The dust-size evolution model}\label{sec:diskmodel}
With the collision model described in the previous section, it is possible to describe the collision outcome between any two particles. In order to study the evolution of the dust in the protoplanetary disk, however, we need to know about the properties of the gas and the sources of relative velocity between the particles. In this section, we describe the disk model used in this paper, along with the dust evolution code of \cite{2010A&A...513A..79B} that has been used together with the new collision model to study the dust-size evolution. A summary of the parameters used for the disk model is given in Table~\ref{tab:diskparams}.

\subsection{The disk model}
\label{subsec:diskmodel}

In this work, we follow the dust-size evolution locally at a distance of 3 AU from the star. To describe the gas distribution over the disk, we use the minimum mass solar nebula (MMSN) \citep{1977Ap&SS..51..153W, 1985prpl.conf.1100H}. This model is based on the current solar system, where the mass of all the planets have been used to predict the minimum total mass that would have been needed to form them. It however excludes the effect of planetary migration and radial drift of dust grains, and the real initial disk profile might have been much different \citep{2007ApJ...671..878D}. It is however useful for comparison with previous collision models. The gas surface density profile of the MMSN is given by
\begin{equation}
	\Sigma_{\rm g}(r) = 1700 \left( \frac{r}{1~{\rm AU} } \right)^{-1.5}~~~[{\rm  g~cm^{-2}}]
\end{equation}
where $r$ is the distance to the central star. At 3 AU, this results in a gas surface density of 330 g cm$^{-2}$, and if we assume an initial dust-to-gas ratio of 0.01, a dust surface density of 3.3 g cm$^{-2}$.

We assume four different sources for the relative velocities between dust grains: Brownian motion, turbulence and azimuthal and radial drift. Since we use a local simulation at a set point, we take into account the relative velocities that arise, but do not allow the particles to move around in the disk. The different sources are discussed briefly below (see \citealp{2010A&A...513A..79B} for a more complete description).

Brownian motion arises from the thermal movement of the particles, and is most effective for the smallest particles. It is dependent on the mass of the particles as follows:
\begin{equation}
	\Delta v_{\rm BM} = \sqrt{ \frac{8 k_{\rm b} T(m_{\rm t} + m_{\rm p}) }{ \pi \cdot  m_{\rm t} m_{\rm p }
	} }
\end{equation}
where $k_{\rm b}$ is Boltzmann's constant, and $T = 115$ K is the gas temperature we assume at 3 AU.

Turbulent motion arises from the particles interaction with the surrounding gas, as it is accelerated by turbulent eddies of different size scales. We use the closed form expressions as derived by \cite{2007A&A...466..413O}. The turbulence strength is given by the $\alpha$ parameter, generally assumed to lie between $10^{-2}$ and $10^{-4}$. The degree at which different particles are affected by the turbulence is given by the Stokes number, denoting how strongly a particle is coupled to the surrounding gas, which for small particles can be written as 
\begin{equation}
	\label{eq:stokes}
	{\rm St} = \frac{\xi a}{\Sigma_{\rm g}} \frac{\pi}{2}~~~~~~~{\rm for}~a < \frac{9}{4} \lambda_{\rm mfp}
\end{equation}
where $\xi = 1.6$ g cm$^{-3}$ is the solid density of the dust grains and $\Sigma_{\rm g}$ the surface density of the gas, and $\lambda_{\rm mfp}$ the mean free path of the gas.

Radial drift gives rise to a relative velocity between particles as they are differently coupled to the surrounding gas \citep{Whipple:3gV1B7CM, 1977MNRAS.180...57W}. This can be written as
\begin{equation}
	\Delta v_{\rm RD} = \left| v_{\rm r}(M_{\rm t}) - v_{\rm r}(M_{\rm p}) \right|
\end{equation}
where the radial velocity of a particle is given by 
\begin{equation}
	\label{eq:raddrift}
	v_{\rm r} = \frac{ v_{\rm g} }{ 1 + {\rm St}^2  } - \frac{ 2 v_{\rm n} }{ {\rm St}  + {\rm St}^{-1}  }
\end{equation}
The first term comes from the drag of the surrounding gas on the particle as the gas migrates radially, and $v_{\rm g}$ is the the velocity of the surrounding gas \citep{1974MNRAS.168..603L}. The second term corresponds to the drift of the particle with respect to the gas. Due to the gas pressure, the gas moves slightly sub-keplerian, and the particle thus experiences a constant headwind which causes it to lose angular momentum and drift inwards. This effect is strongly related to the coupling between the particle and the gas. $v_{\rm n}$ represents the maximum drift velocity, and is derived by \cite{1977MNRAS.180...57W} as
\begin{equation}
	v_{\rm n} = - \frac{ \frac{ \partial P_{\rm g} }{ \partial r} }{ 2 \rho_{\rm g} \Omega_{\rm k} } = 3.9 \cdot 10^3~~~{\rm cm~s}^{-1}
\end{equation}
where $\frac{ \partial P_{\rm g} }{ \partial r}$ the gas pressure gradient, $\rho_{\rm g}$ the gas density and $\Omega_{\rm k}$ the Kepler frequency.

Azimuthal relative velocities work in a similar fashion to radial drift, and arise from gas drag in the azimuthal direction. The relative azimuthal velocity can be written as
\begin{equation}
	\Delta v_\varphi = \left| v_{\rm n} \cdot \left( \frac{1}{1 + {\rm St}_{\rm t}^2} - \frac{1}{1 + {\rm St}_{\rm p}^2} \right) \right|
\end{equation}
\begin{figure}[t]
\centering
\resizebox{0.9\hsize}{!}{\includegraphics{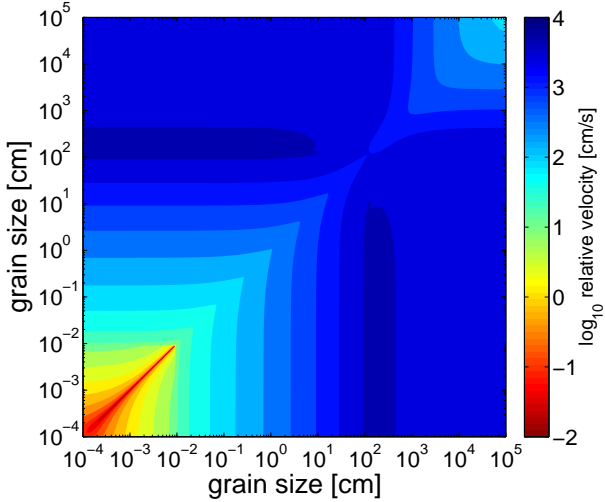} }
\caption{The relative velocities for each particle pair calculated from the four sources described in Sect.~\ref{subsec:diskmodel} using the parameters given in Table~\ref{tab:diskparams}.}
\label{fig:relvelocities}
\end{figure}
\begin{table}
\caption{Disk model parameters used in the simulations.}
\begin{tabular}{ l l l l }
\hline
\hline
Parameter & Symbol & Value & Unit \\
\hline
distance to star 	& $r$ 			& 3 			& AU \\
gas surface density 	& $\Sigma_{\rm g}$ 	& 330 		& g cm$^{-2}$ \\
dust surface density & $\Sigma_{\rm d}$ 	& 3.3 		& g cm$^{-2}$ \\
gas temperature 	& $T$ 			& 115 		& K \\
turbulence parameter & $\alpha$		& $10^{-3}$	& - \\
maximum drift velocity & $v_{\rm n}$		& $ 3.9 \cdot 10^3$ & cm s$^{-1}$ \\
sound speed		& $c_{\rm s}$		&  $6.4 \cdot 10^4$ & cm s$^{-1}$ \\
solid density of dust grains & $\xi$ 			& 1.6 		& g cm$^{-3}$ \\
\hline
\end{tabular}
\label{tab:diskparams}
\end{table}
In Fig.~\ref{fig:relvelocities}, we plot the resulting relative velocity field between each particle pair taking into account the four sources described above. For particles smaller than $\sim 10$ $\mu$m, Brownian motion is the dominant source of relative velocity, causing velocities on the order of mm$^{-1}$. At larger sizes, turbulence becomes important, and velocities quickly increase to $\sim 1$ m s$^{-1}$. As can be seen in Eq.~\ref{eq:raddrift}, the radial drift is largest for particles with a Stokes number of 1, which at 3 AU corresponds to around 30 cm in size. At roughly this size, due to the combined effect of radial and azimuthal drift, the particles collide with the smaller particles at velocities around 50 m s$^{-1}$, which then decreases to 40 m s$^{-1}$ as the particles grow larger and the radial drift decreases.

\subsection{The Smoluchowski equation}
\label{subsec:solver}

The dust grain number density $n(m, r, z)$ is a function of the grain mass, $m$, the distance to the star, $r$, and the height above the mid-plane, $z$, and describes the number of particles per unit volume per unit mass. The total dust density can therefore at a point $(r,z)$ be written as
\begin{equation}
	\rho(r,z) = \int_0^\infty{n(m,r,z) \cdot m~} {\rm d}m
\end{equation}
and the change in number density with respect to time can then be given by the Smoluchowski equation as
\begin{align}
	\frac{\partial}{\partial t}n(m,r,z) = \int \int_0^\infty & M(m,m',m'',r,z) \\ \nonumber \times~ & n(m',r,z) n(m'',r,z)~{\rm d}m'{\rm d}m''
\end{align}
where $M(m,m',m'',r,z)$ is called the kernel, and describes how the mass $m$ is distributed after an interaction between particles $m'$ and $m''$. This distribution is determined by the use of a collision model like the one developed in this paper and is described in Sect.~\ref{sec:implementation}. See \cite{2010A&A...513A..79B} on how one constructs $M$ out of a collision model.

In the code implementation, the density distribution is discretized over logarithmically spaced mass bins. The resulting mass(es) of a collision between two particles will generally not coincide with one specific mass bin. In order to solve this, the resulting mass is therefore divided between the two neighbouring mass bins by using the Podolak algorithm described in detail by \cite{2008A&A...480..859B}.

In order to solve the above equation and track the size-evolution of the dust grains, we use an implicit scheme developed by \cite{2008A&A...480..859B} and \cite{2010A&A...513A..79B}. This scheme allows for longer timesteps and therefore shorter execution times.

\section{Results}\label{sec:results}
We have performed local simulations of the dust-size evolution using the collision model described in Sect.~\ref{sec:implementation} and the evolution code briefly described in Sect.~\ref{sec:diskmodel}. In this section, we discuss the outcome of the new collision model and compare it to previous models. We also show the results of the simulations and compare the growth of the large particles to a simple analytical model.

\subsection{The collision outcome space}
\label{subsec:colloutcome}

\begin{figure}[t]
\centering
\resizebox{1.0\hsize}{!}{\includegraphics{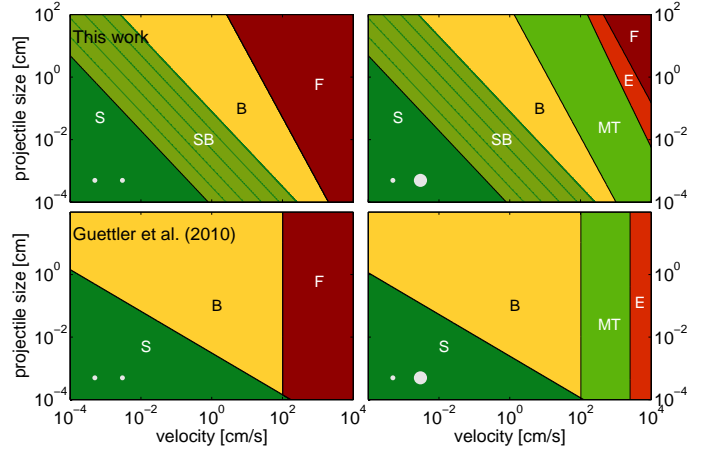}}
\caption{Comparison between the new collision model (top) and the model of \cite{2010A&A...513A..56G} (bottom). The left and right panels show the outcome for equal- and different-sized collisions, respectively. Green regions mark collisions which are growth-positive for the target, yellow marks growth-neutral and red marks growth-negative. 'S' marks sticking, 'SB' the sticking to bouncing transition, 'B' bouncing, 'MT' net mass transfer, 'E' net erosion and 'F', fragmentation. In the transition region, the green parallel lines each mark a decrease in sticking probability by 20\%.}
\label{fig:comparison}
\end{figure}

With the new collision model, we can determine the outcome after a collision between two particles of arbitrary masses and velocities. In the upper panels of Fig.~\ref{fig:comparison}, we plot the collision outcome as a function of projectile size and collision velocity for two different mass-ratios. This can be compared to the outcome of \cite{2010A&A...513A..56G} for compact particles shown in the bottom panels. It can here be noted that our model naturally describes the transition between the two extreme cases of equal-sized and different-sized collisions, while G\"uttler et al. had to define a critical mass ratio to distinguish between the two regimes. The upper right panel in the figure thus only gives a single snapshot in this transition.

In the left panels, the two particles are of equal size, and the models produce very comparable results. In the new model, the sticking region has been enlarged with the inclusion of a transition region where both sticking and bouncing is possible. In the fragmentation region, the mass-dependent fragmentation threshold has decreased the velocity needed to fragment larger particles, and increased the velocity needed for the smallest particles. The end outcome is that the width of the bouncing region has decreased significantly.

In the right panels, the target has a mass 1000 times the mass of the projectile. Some important differences can here be seen in the fragmentation regime. We can first of all notice the new natural transition from growth to erosion which is obtained from the balance between growth from fragmentation with mass transfer and erosion from cratering (Eq.~\ref{eq:masschange}). At this mass-ratio, erosion transitions quickly into complete fragmentation. If the mass-ratio was to be increased further, the fragmentation region decreases and is replaced with erosion.

As long as the projectile is fragmenting, velocities below the erosion threshold always lead to growth, and a cm-sized projectile can initiate mass-transfer already at about $20$ cm s$^{-1}$ (which is exactly the result of \citealp{2011ApJ...736...34B}). At $\Delta v = 10$ m s$^{-1}$, projectiles smaller than around 1 cm are required for growth. The maximum projectile size decreases with velocity as the erosion grows stronger, and at $\Delta v = 50$ m s$^{-1}$, growth is only possible with projectiles smaller than $100$ $\mu$m.

Overall, we predict more fragmentation and cratering than in the previous model of \cite{2010A&A...513A..56G}. However, one very important change is that growth via fragmentation with mass transfer is now possible at higher velocities than the 20 m s$^{-1}$ that was the previously predicted threshold, and provided that the projectile is small enough, even a collision at 50 m s$^{-1}$ that was predicted in the disk model can lead to growth of the target (which was a direct conclusion of \citealp{2009MNRAS.393.1584T}).

Sticking collisions are also possible at larger sizes, and growth-positive mass transfer works at much lower velocities than the previously assumed 1 m s$^{-1}$. Even if the bouncing region shrinks in size, we will see that it will not be enough to remove the bouncing barrier. If we insert a particle above the bouncing barrier, however, the required relative velocity for it to interact beneficially with the particles below the bouncing barrier has been decreased. These two updates turn out to be quite important, as is discussed in more detail in Sect.~\ref{subsec:dustev}.

\begin{figure}[t]
\centering
\resizebox{0.9\hsize}{!}{\includegraphics{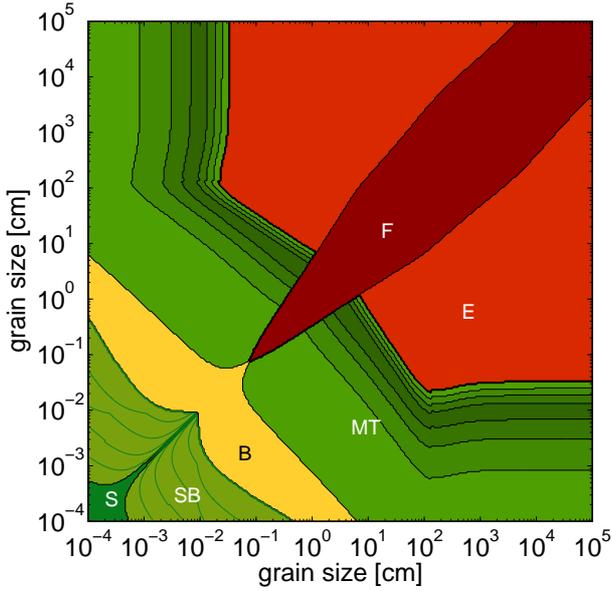}}
\caption{The collision outcome for all pairs of particles with the relative velocity field calculated in Fig.~\ref{fig:relvelocities} and with the same labels and color code as in Fig.~\ref{fig:comparison}. Also included is the net mass transfer efficiency, given in intervals of 4\%. }
\label{fig:collisionspace}
\end{figure}

\begin{figure}[h]
\centering
\resizebox{1\hsize}{!}{\includegraphics{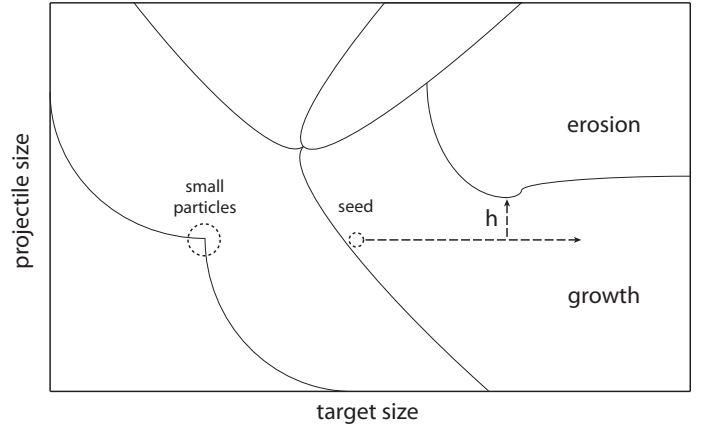}}
\caption{A zoomed-in sketch of the collision outcome space shown in Fig.~\ref{fig:collisionspace}. The dashed horizontal line shows the interaction path that the seed will experience during its growth. The $h$ parameter illustrates the minimum distance between the interaction path and the erosive region. A positive $h$ means that the boulder/small particle interactions will always be growth-positive, and a negative $h$ means that the growth will at some point be stopped by erosion.}
\label{fig:outcomezoom}
\end{figure}

The collision outcome for the new model depends on the mass of both the projectile and the target, and in the current disk model, we use only the average relative velocity between each particle pair. This means that a collision between a given pair will in the evolution model always result in the same outcome, and it would therefore be instructive for us to plot the outcome in the particle size-size space. In Fig.~\ref{fig:collisionspace}, we have used the relative velocity field calculated in Sect.~\ref{subsec:diskmodel} at a distance of 3 AU to determine the outcome for each collision pair.

In this figure, the bouncing barrier is clearly visible. Dust grains of sizes $100-800$ $\mu$m that interact with smaller particles will because of too large collision velocities bounce if the particle is not smaller than $10$ $\mu$m. In this case, a small number of collisions will lead to sticking, but in order to pass the wide bouncing region, a grain would need to experience $10^9$ such sticking collisions. The small particles will however be able to coagulate to $100$ $\mu$m themselves, making growth through the bouncing barrier very difficult.

Collisions between two equal-sized particles larger than 1 mm will result in destructive fragmentation, but depending on what it collides with, a 1 mm-sized particle can also be involved in sticking, bouncing, mass transfer and erosive collisions. Because of the fragmentation with mass transfer effect, a meter-sized boulder can grow in collisions if its collision partner is of the right size, in this case smaller than $200$ $\mu$m. As we can see in this plot, the key to growing large bodies is therefore to sweep up smaller particles faster than they get eroded or fragmented by similar-sized collisions.

From Fig.~\ref{fig:collisionspace}, we can already without performing any simulations see that a cm-sized particle would be capable of growing to large sizes if it collides with the right projectiles. The important parameter needed to determine this is illustrated in Fig.~\ref{fig:outcomezoom}, showing a sketch of a part of the collision outcome plot. Because of the bouncing region, most of the particles will be found in the region marked in the figure. A boulder needs to interact beneficially with these bouncing particles in order to grow, so the horizontal interaction path needs to be at all times in the growth-positive mass-transfer region. This can be illustrated with the $h$ parameter, which gives the minimum difference between the interaction path and the erosive region. If $h$ is positive, the boulder will always interact beneficially with the bouncing particles, but if $h$ for some reason would become negative, the growth of the boulder would stop.

We can now point out the interesting effect turbulence has on the collision outcome. For particles of sizes between $10$ $\mu$m and 10 cm, turbulence is the dominant velocity source. If the relative velocity is increased in this regime, the bouncing barrier will be pushed down to smaller sizes. The larger particles are however not as much affected by a stronger turbulence, as these sizes are also affected by radial and azimuthal drift. This means that the $h$ parameter will remain constant or possibly even increase with a stronger turbulence. A strong turbulence might therefore even be beneficial for this mode of growth, as the larger particles will now interact with generally smaller particles, which we from Fig.~\ref{fig:craterbalance} know is beneficial for the mass transfer effect. Because of this, even if the boulders due to strong turbulence are experiencing relative velocities of $\sim$100 m s$^{-1}$, they can grow in interactions with the small particles at the bouncing barrier, as these have correspondingly decreased in size.

\subsection{The dust-size evolution}
\label{subsec:dustev}

\begin{figure}
\centering
\resizebox{1.0\hsize}{!}{\includegraphics{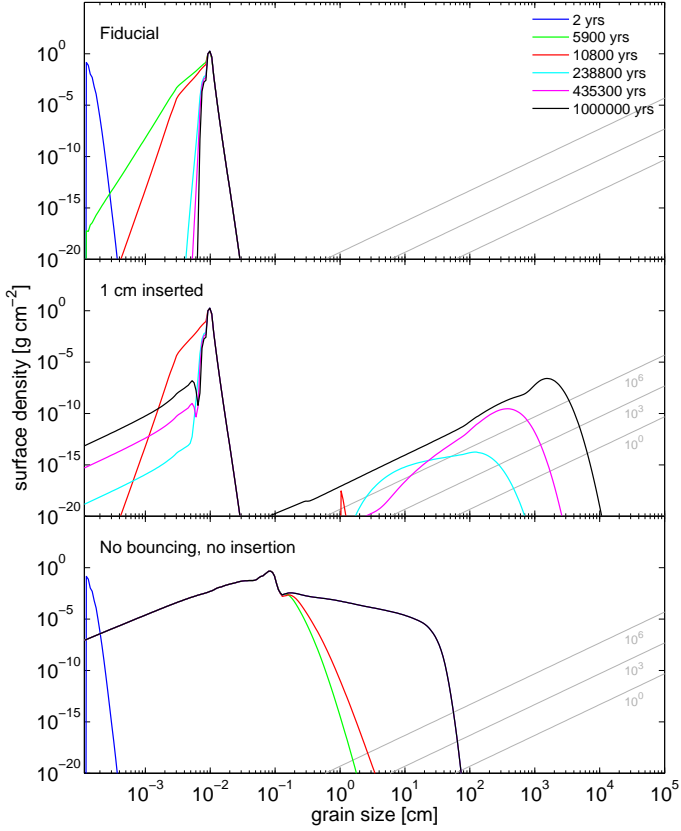}}
\caption{The surface density evolution of the dust population for three different simulations at a local simulation at 3~AU. The grey diagonal lines correspond to the required surface densities for a total number of particles of $1$, $10^3$ and $10^6$ in an annulus of thickness 0.1 AU. In the \textit{upper panel}, all particles initially have a size of $10^{-4}$ cm, and snapshots are taken between 2 and $10^6$ years. In the \textit{middle panel}, we have run the same simulation, but after 10,800 years, a small number of 1 cm-particles have artificially been inserted. In the \textit{lower panel}, the bouncing barrier has been replaced with sticking, allowing the particles to freely coagulate to larger sizes.}
\label{fig:numdens}
\end{figure}

We performed simulations using the new collision model together with the local version of the \cite{2010A&A...513A..79B} continuum dust-size evolution code. In Fig.~\ref{fig:numdens}, the mass distribution of the particle sizes is given at different timesteps for the three different experiments discussed in detail below.

\subsubsection{Growth up to the bouncing barrier}

In the fiducial case presented in the top panel, the simulations are initiated with all dust made up of $\mu$m-sized monomers. At these small sizes, the relative velocity is driven by Brownian motion, and as the particles collide with each other, they stick together and form larger aggregates. This leads to a rapid coagulation phase where the aggregates grow to 100 $\mu$m in around 1000 years. At this point, the particles have grown large enough to become affected by the turbulence which quickly increases the relative velocities. As we predicted in Fig.~\ref{fig:collisionspace}, the bouncing region is too wide to be passed over, and the growth halts at the bouncing barrier.

At this stage, the only particles that can grow are the smaller ones, and as time proceeds, more and more particles get trapped at the bouncing barrier. This causes the number of small particles to keep on decreasing, leading to a continuously narrowing size-distribution. After $10^5$ years, virtually all particles can be found at 100 $\mu$m, with very steep tails between 60 and 300 $\mu$m. If nothing else is done, this is how the dust evolution ends. The bouncing barrier efficiently prevents any further growth, and all particles remain small.

\begin{figure*}[t]
\centering
	\includegraphics[width=15cm]{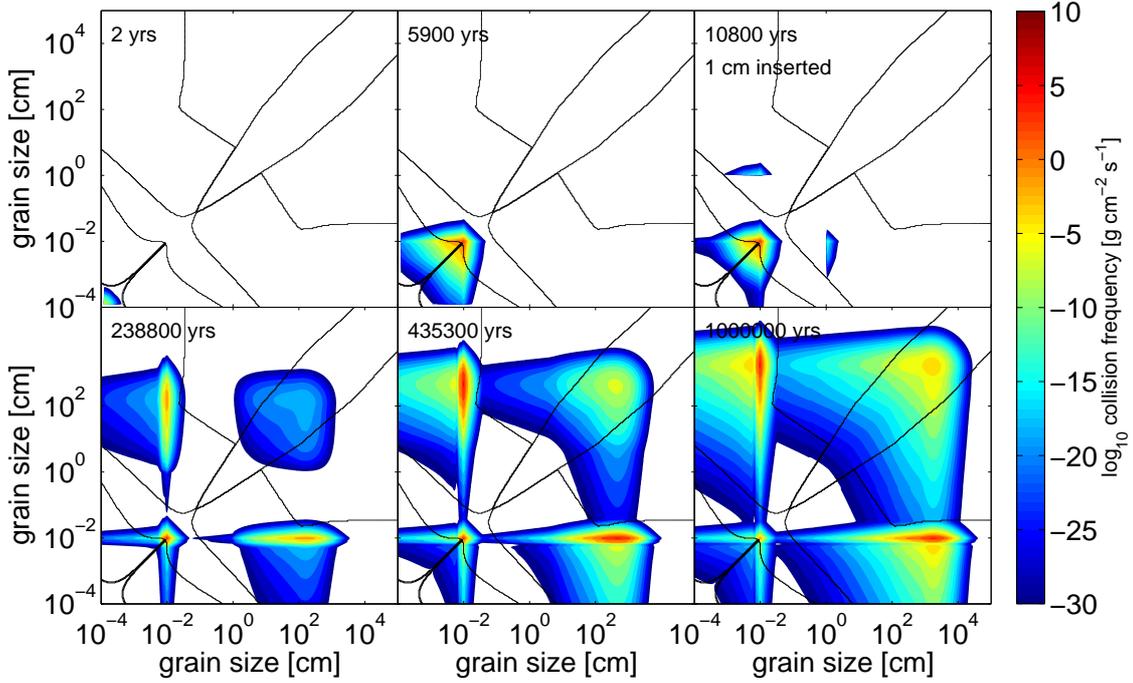}
\caption{The collision frequency map for the scenario where 1 cm-particles are artificially inserted at $t = 10,800$ years. The interaction frequency is plotted for each particle pair at six different timesteps plotted on top of the collision outcome space of Fig.~\ref{fig:collisionspace}. This makes it possible to identify the dominating interaction for each particle size. Note the always high peak of interactions with particles stuck below the bouncing barrier at 1 mm in size. As the large particles grow, they also sometimes collide among themselves, producing a tail of particles capable of also sweeping up the bouncing particles. This causes an increase in both mass and number for the large particles, and a continuous widening of the size-distribution.}
\label{fig:collfreqmap}
\end{figure*}

\subsubsection{A seeding experiment}
\label{subsec:seeding}

In order to investigate the potential of the mass transfer effect, we make an experiment where a very small number (i.e. $10^{-18}$ of the total mass) of "seeds" are artificially inserted in the form of 1 cm-particles. As can be seen in Fig.~\ref{fig:collisionspace}, the interaction between 1 cm and 100 $\mu$m-particles results in mass transfer and growth of the larger particle, and we expect the inserted particles to be able to grow. The seeds are inserted at a single time $t = 10,800$ years, when the particles have reached the semi-stable state at the bouncing barrier, and the result can be seen in the middle panel of Fig.~\ref{fig:numdens}. Exactly how the seeds are formed will not be discussed in this paper, but given the small number of seeds required, stochastic effects, small variations in local disk conditions or grain composition and/or properties might suffice to produce them. Some other possibilities are briefly discussed in Sect.~\ref{subsec:bouncingbarrier}.

In order to better understand the complex interaction between all the particles in this experiment that now follows, we introduce the collision frequency plot given in Fig.~\ref{fig:collfreqmap}. This shows the collision frequency between each particle pair plotted on top of the collision outcome map of Fig.~\ref{fig:collisionspace}, making it possible for any given time to identify the dominating collision type for a given particle size.

The first two snapshots in the collision frequency plot are taken after 2 and 5900 years, and are identical to the fiducial case discussed earlier. At the bouncing barrier, we can see some interaction between the ~200 $\mu$m particles and the smallest particles that do lead to growth, but the frequency is much too small to have any significant effect.

After $10,800$ years, the 1 cm seeds are inserted, and they grow to larger boulders by sweeping up the small particles trapped at the bouncing barrier. As the boulders grow, one can after $\sim$200,000 years see a tail of particles with intermediate sizes appear behind them. These are formed by the rare collisions between the large boulders, and from a single event, two large bodies have been multiplied to a myriad of fragments also capable of sweeping up the particles at the bouncing barrier. This effect causes the population of boulders to not only grow in total mass, but also in number, which causes a steady and significant increase in similar-sized fragmentation.

It can also be noted how the vertical distribution of dust around the midplane affects this stage of the evolution. Even if there is a huge amount of particles trapped at the bouncing barrier, they are so small that many of them are pushed out from the midplane due to turbulent mixing. The boulders are however so large that they have decoupled significantly from the gas, and are therefore mostly trapped in the midplane. This causes the sweep-up rate of the mm-sized particles by the boulders to be distinctly lower than without a vertical structure, and also the internal collisions between the boulders to be relatively more common.

The smallest particles that are produced by global fragmentation and erosion mainly experience two different interactions. In the early stages, the smallest fragments are generally being swept up by the 100 $\mu$m-sized particles stuck at the bouncing barrier, since these particles dominate completely both in number and mass. They have therefore never any time to coagulate to larger sizes themselves, but instead aid in the growth of the bouncing particles, which are in turn swept up by the boulders. At later stages, as the boulders become more numerous, it is also possible that the smallest fragments are swept up directly by the boulders. If this growth continues even longer, the two effects become equally efficient, and even later, the boulders will start dominating in the sweep-up. Regardless of what sizes the small fragments interact with, in the end, they are still beneficial for the growth of the boulders.

In the end, a number of 10-70 m boulders have managed to form, and the amount of total mass in the large particles have increased by the huge factor of $10^{12}$ from what was initially inserted into the system, even though the total boulder mass is still very small compared to the total dust mass. We find that the limiting case for the growth at this point is not so much erosion or fragmentation as it is the growth timescale (see also \citealp{2008A&A...486..597J}). If the simulation is kept running for longer than $10^6$ years, the boulders can keep on growing and several hundred-meter boulders can form. In other places in the disk with higher dust densities and relative velocities, larger boulders will be able to form on the same time-scale.

Growth by sweep-up gives an explanation of how the collision part of the growth barrier can be circumvented, but we have in these simulations disregarded the effect of the orbital decay from gas drag. The growth timescales in Fig.~\ref{fig:numdens} exceed by several orders of magnitude the lifetime of meter-sized bodies subject to radial migration. In order to survive, the bodies need to either form on a timescale very much smaller than observed in our simulations, which we find unlikely, or there need to exist some effect which prevents the orbital decay over an extended period of time \citep{1995A&A...295L...1B, 1997Icar..128..213K, 2007A&A...469.1169B, 2011arXiv1112.2349P}.

\subsubsection{Removing the bouncing barrier}

In order to illustrate the importance of the bouncing barrier, we make an experiment where all the bouncing collisions are removed and replaced with sticking. There will therefore be nothing that prevents the coagulation phase from continuing also to larger sizes. The result of this simulation is shown in the lower panel of Fig.~\ref{fig:numdens}.

This allows the particles to grow unhindered until they reach about 1 mm in size. At this point, they will start to fragment between themselves, as we can see in Fig.~\ref{fig:collisionspace}. Since most of the dust can be found at this size, heavy fragmentation will occur, and a cascade of smaller particles are produced. These small particles grow up to larger sizes again, where most of them again fragment. From the collision outcome plot, we can however see that some particles can be lucky, and instead only sweep up smaller particles via mass transfer. They therefore avoid fragmentation, and keep on growing, and form the distribution tail extending from 1 mm and upwards.

At a size of roughly 1 meter, the only way to grow is by colliding with $100$ $\mu$m or smaller particles. But without the bouncing barrier, most of the mass is instead found in 1 mm particles. This means that almost all of the interactions become erosive, and the growth halts.

In the above example, we saw that increasing the stickiness of particles in the end actually \textit{prevented} the growth of large boulders. We can from this work draw the conclusion that the bouncing barrier might even be beneficial for the formation of planetesimals. By hindering the growth of the many, it makes it possible for the growth of a lucky few.

\subsubsection{The effect of turbulence}

\begin{figure}
\centering
\resizebox{1.0\hsize}{!}{\includegraphics{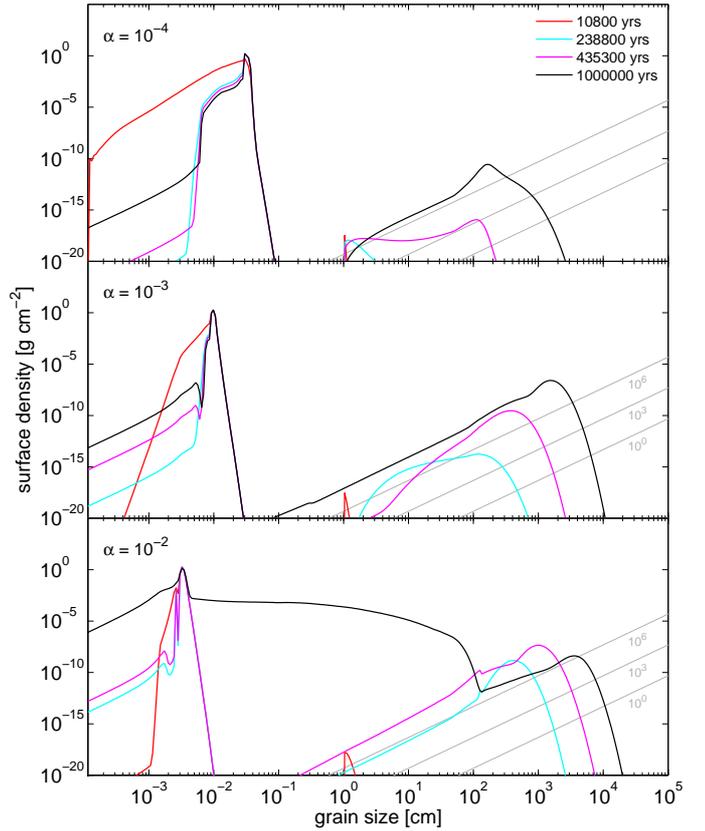}}
\caption{The surface density evolution of the dust population for three different turbulence strengths $(\alpha = 10^{-4}, 10^{-3}, 10^{-2})$ at 3~AU, with seeds inserted artificially after $t = 10,800$ years.}
\label{fig:numdens_alpha}
\end{figure}

We have so far studied only one single turbulence parameter of $\alpha = 10^{-3}$. This value is however uncertain for nebular models, and in order to investigate the dust evolution for different velocity fields, we also study the cases of $\alpha = 10^{-4}$ and $\alpha = 10^{-2}$. The latter represents a strong turbulence which is dominating completely over the azimuthal and radial drifts and results in relative velocities up to 100 m s$^{-1}$. In the former case, the turbulence is weak, and contributes very little to the velocities of the larger particles. In Fig.~\ref{fig:numdens_alpha}, we plot the resulting size evolutions for the three turbulence parameters but with otherwise identical initial conditions. The middle panel is here the same as the middle panel of Fig.~\ref{fig:numdens} and is included for reference.

It can first of all be noted that growth to larger sizes is possible even in the case of a very high turbulence, as we predicted using Fig.~\ref{fig:outcomezoom}. This is due to the decrease in size of the bouncing particles as the turbulence increases. The smaller impacting projectiles can therefore cause the boulder to grow also at higher velocities, as seen in Fig.~\ref{fig:craterbalance}.

The growth timescale is affected by the turbulence in several different ways. Firstly, it increases the relative velocities between the particles, leading to higher collision frequencies and therefore to more rapid growth. Increased impact velocities also increases the net mass transfer efficiency, which in Fig.~\ref{fig:craterbalance} is seen to be particularly low for velocities below 10~m~s$^{-1}$. An increased turbulence also mixes small particles further out from the midplane, which decreases their midplane densities where the largest boulders gather, lowering the growth rate. From Fig.~\ref{fig:numdens_alpha}, it is clear that the growth-positive effects are dominating, leading to enhanced growth with increased turbulence. In the low turbulence case, the growth is especially slow for the sizes between 1 cm and 100 cm. In this regime, the relative velocities are very low since the contributions from the azimuthal and radial drift are small, causing very low net mass accretion efficiencies during the sweep-up growth.

In the $\alpha = 10^{-2}$ case, we can in the final snapshot see that a separate peak has appeared for the largest boulders. It forms when the boulders have fragmented so much between themselves that the large intermediate-sized fragments are at number densities roughly equal to the boulders, which in this case happens after $\sim 600,000$~years. At this point, the boulders are significantly eroded and fragmented, which creates even more intermediate-sized particles capable of yet more fragmentation. This results in a fragmentation cascade, and a rapid flattening of the size-distribution. This does however not cause all the large particle to be destroyed, and those that survive can continue to grow by sweeping up the small bouncing particles that still dominate both in number and total mass. This effect occurs also for the cases of weaker turbulence if the simulations run long enough, but is never severe enough to cause a complete halt of the growth.

We have have here shown that growth can proceed also in regions of high turbulence. An MRI turbulent disk might however also create an additional velocity source, where local gas density fluctuations can excite the orbital eccentricities of the planetesimals. \cite{2008ApJ...686.1292I} found that this can lead to velocities beyond break-up even for small planetesimals, but only studied collisions between equal-sized bodies. In our model where only the material strength of the body is included, a collision between two large and similar-sized bodies will be destructive even in regions of very low turbulence, so in this regime, it will affect our conclusions very little. Excited planetesimal orbits will however also increase the impact velocities of the small particles, which could pose a problem. In the previous cases, growth has been possible  in regions of high turbulence because the sizes of the bouncing particles have simultaneously been suppressed by the turbulence, but this is not necessarily the case in the scenario studied by Ida et al. This might cause an increased erosion of the growing planetesimals, but investigating this further is outside the scope of this paper.

\subsection{A simple analytical model}

To verify what we see in the simulations, where growth is caused by the sweep-up of small particles by a few lucky seeds, we will make a simple analytical growth model in a similar fashion to \cite{2008A&A...486..597J} and \cite{2010ApJ...724.1153X}. We consider a scenario where a single large body of mass $m$ is moving around in a sea of small particles of mass $m_{\rm s}$ stuck at the bouncing barrier. The growth of the large body can then be described by
\begin{equation}
	\label{eq:dmdt}
	\frac{ {\rm d}m }{{\rm d}t}= \sigma \Delta v \cdot \epsilon_{\rm net} n_{\rm s} m_{\rm s}
\end{equation}
where $\sigma = \pi \left( a_{\rm s} + a \right)^2 \sim \pi a^2$ is the collisional cross-section, $\Delta v$ the relative velocity, $n_{\rm s}$ the number density of the small particles and $\epsilon_{\rm net}$ the average net mass gain efficiency of a collision. The change in size $a$ with respect to mass can be written as ${\rm d}m = 4 \pi \xi a^2~{\rm d}a$, where $\xi = 1.6$ g cm$^{-3}$ is the internal density of the large body. We can now write 
\begin{equation}
	\label{eq:dadt}
	\frac{ {\rm d}a }{ {\rm d}t } = \frac{1}{4} \frac{ \rho_{\rm s} }{ \xi } \epsilon_{\rm net} \Delta v
\end{equation}
where $\rho_{\rm s} = n_{\rm s} m_{\rm s}$ is the mass density of the small particles. We will for now assume that the relative velocity is caused only by turbulence, and can therefore for small particles be estimated by \citep{1993prpl.conf.1031W,2001ApJ...546..496C}
\begin{equation}
	\label{eq:dv}
	\Delta v = \sqrt{ \frac{9}{2} \alpha {\rm St} } \cdot c_{\rm s}
\end{equation}
where $\alpha = 10^{-3}$ is the turbulence parameter, St the Stokes number for the boulder given by Eq.~\ref{eq:stokes} and $c_{\rm s}$ the sound speed in the gas given by
\begin{equation}
	c_{\rm s} = \sqrt{ \frac{ k_{\rm b} T }{ \mu m_p } }
\end{equation}
where $k_{\rm b}$ is Boltzmann's constant, $T = 115$ K is the temperature of the gas, $\mu = 2.3$ the mean molecular weight and $m_p$ the mass of a proton.

\begin{figure}[t]
\centering
\resizebox{1.0\hsize}{!}{\includegraphics{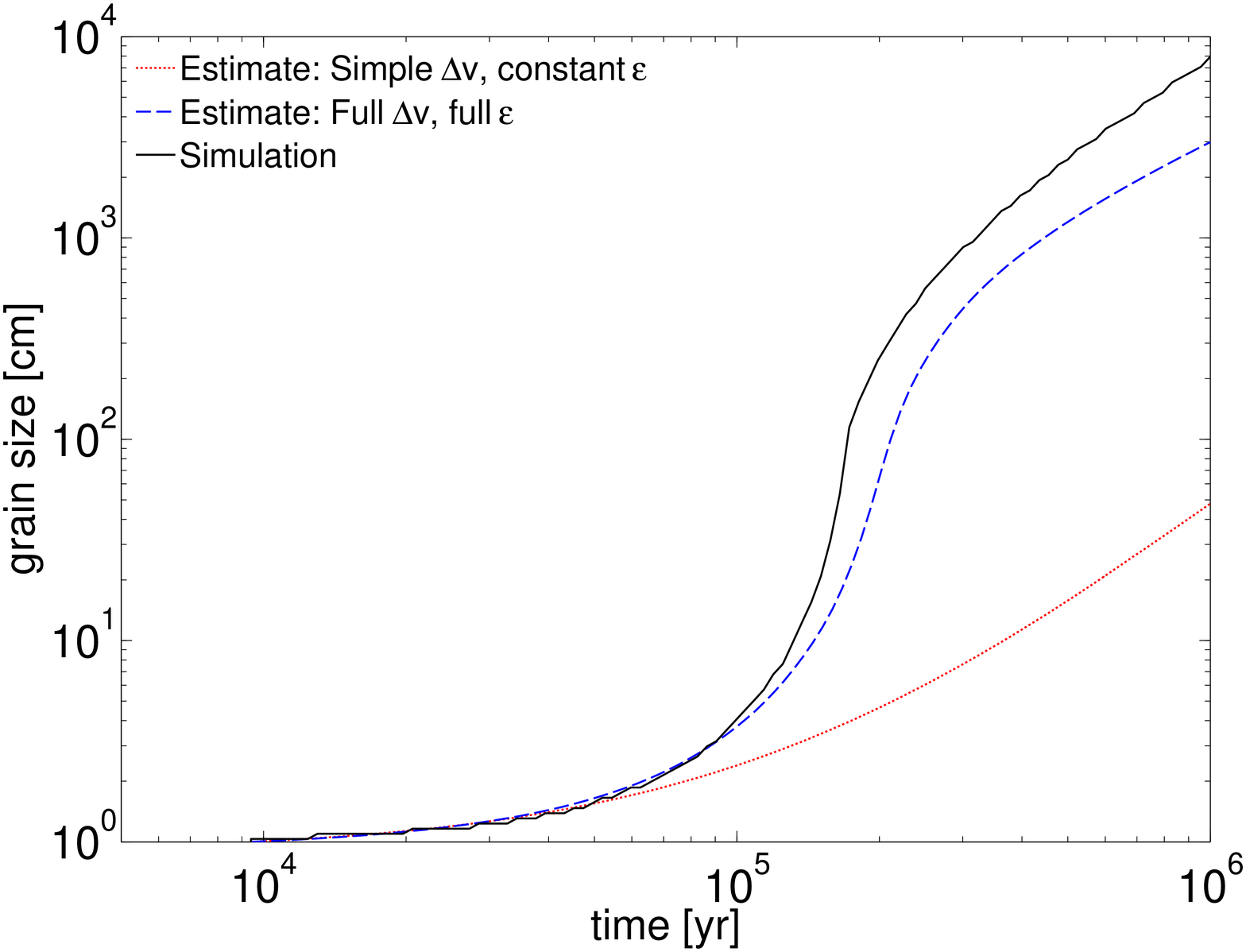}}
\caption{The growth of a boulder in the simulations (black) compared to two growth estimates. The red line corresponds to the simplified analytical estimate, and the blue line to the estimate using the full relative velocity and accretion efficiency prescriptions.}
\label{fig:analyticalcomp}
\end{figure}

We assume $\rho_{\rm s}$ to be constant and unaffected by the sweep-up of the large boulder. When the boulder has grown large enough, it will be very little affected by the vertical mixing from turbulence, and will mostly be found near the midplane. We therefore take $\rho_{\rm s} = 10^{-12}$ g cm$^{-3}$ to be the midplane density assuming a Gaussian vertical profile with a surface density $\Sigma_{\rm s} = 3.3$ g cm$^{-2}$ and that all the of the dust mass can be found in the small particles. Eq.~\ref{eq:dadt} can then be solved analytically:
\begin{equation}
	a(t) = \left( \frac{1}{8} \epsilon_{\rm net} \frac{ \rho_{\rm s} }{ \xi } \sqrt{ \frac{9}{4} \frac{\xi}{\Sigma_{\rm g}} \pi \alpha } \cdot c_{\rm s} \left( t - t_0 \right) + \sqrt{a_0} \right)^2
\end{equation}
In Fig.~\ref{fig:analyticalcomp}, we plot the above analytical solution to the boulder growth assuming $\epsilon_{\rm net} = 0.006$ which is the expected efficiency for the interaction between a 1 cm and a 1 mm body. We assume the other parameters identical to the simulations. This is compared to the observed growth of the largest particle from the seeding simulations in Sect.~\ref{subsec:seeding} using the full collision model. As can be seen, our analytical solution compares well to the growth rate of the simulations for the first stage of growth, and then at later stages differ greatly. 
This is mainly due to two reasons. Assuming a constant $\epsilon_{\rm net}$ means that it at larger sizes will be underestimated, as it in the full model depends on the collision velocity and varies between 0 and 0.15. The relative velocity prescription in Eq.~\ref{eq:dv} only takes into account the turbulence, and is also not valid for large particles. In order to take this into account, we use the full net growth efficiency from Eq.~\ref{eq:masschange} and all four velocity sources discussed in Sect.~\ref{sec:diskmodel}. We then solve Eq.~\ref{eq:dadt}, and plot the resulting solution in Fig.~\ref{fig:analyticalcomp}. The new solution and the simulation results now agree very well, only to differ slightly at the largest particle sizes.

We find that this enhanced growth is triggered by the existence of the tail of intermediate-sized particles that form from fragmenting events between large bodies. The largest boulders found in the simulations are those that manage to grow by avoiding interactions with other large bodies, and if they manage to interact beneficially with some of the intermediate-sized fragments, the growth rate is enhanced.

%
%
%
%
%
%
%
%
%

\subsection{Forming the first seeds}
\label{subsec:bouncingbarrier}

The width of the bouncing barrier has with recent sticking and fragmentation experiments decreased, but is still solid enough to prevent particles from growing through it. As previously discussed, this might even be positive for the planetesimal formation, as it prevents too many large bodies from forming, keeping them from fragmenting and eroding between themselves. It is however necessary for some cm-sized seeds to be formed that can initiate the sweep-up.

We have in this work found that the bouncing barrier can be circumvented, and growth via mass transfer initiated, even if only a very small number of cm-sized particles is introduced to the system. We have in this paper not investigated in detail how this might happen, but there are many possibilities. Larger particles may form outside the snow-line mixed with ices, and drift inwards to a region where sweep-up becomes possible. Calcium- and alumunum-rich inclusions (CAIs) are cm-sized particles that are believed to have been formed early near the Sun and transported outwards in the disk \citep{2009Icar..200..655C}, and could also make up the first generation of seeds.

It is also possible for some lucky particles to grow through the barrier simply by interacting with enough smaller particles, as can be seen in the collision outcome plot in Fig.~\ref{fig:collisionspace}. In order for this to happen, however, it is necessary to have a wide enough size-distribution so that there is enough of the small particles that the lucky particles can interact with. This is difficult, it turns out, as we have in these simulations found that the particle size-distribution will quickly narrow as all the small dust coagulate up to the bouncing barrier. If extra mass is continuously introduced to the system the first 10,000 years or so, for example due to nebular infall of the collapsing protostellar cloud, the particle size-distribution could be wide enough for some seeds to be formed. \cite{2011arXiv1105.3897B} also found in laboratory experiments that chondrules and dust aggregates tend to stick at higher velocities than collisions between two dust aggregates, so that early chondrules could grow where dust could not.

Another possibility is the introduction of a velocity distribution for each particle-size. Most work with the dust-size evolution consider only the average relative velocity between two particles, but in reality, some particles will collide also at much higher and lower velocities. If some particles are lucky enough to only collide with others at low velocities, they might stick together where they would otherwise only bounce, and in that grow large enough to initiate sweep-up.

\section{Discussion and Conclusions}\label{sec:discussion}
In order to explore the possibility of growth to planetesimal sizes by dust collisions, we have implemented a new collision model which is motivated both physically and experimentally. It is streamlined to work with continuum codes, while still able to take into account all important collision types that have been identified in previous work. Using it with continuum dust evolution codes makes it possible to resolve all dust grains regardless of their numbers, something which we have found to be essential for the study of growth of dust grains above cm-sizes.

Even though collisions between large similar-sized dust grains generally lead to fragmentation, this is not necessarily true for larger mass-ratios if the projectile is small enough. As shown in laboratory experiments, if the projectile is smaller than 0.1-1 mm in size, fragmentation with mass transfer can cause growth of the target at impact velocities as high as 60 m/s. In order for dust grains to grow larger than cm-sizes via collisions, the number of large particles therefore has to be very small to avoid destructive fragmentation among themselves, and most of the dust mass kept at small sizes.

We have found in our simulations that direct growth of planetesimals via dust collisions still is a possibility. We find that the bouncing barrier introduced by \cite{2010A&A...513A..57Z} might be beneficial or even vital for the planetesimal formation, as it provides a natural way to keep most of the dust population small. This small dust makes ideal material for the sweep-up process if larger bodies manage to form. By artificially inserting a few 1 cm-sized seeds in our simulations, we find that they can sweep up the small dust via fragmentation with mass transfer and grow to $\sim$100 meters in size on a timescale of 1 Myr. This opens up exciting new possibilities that need to be taken into account when studying dust growth in protoplanetary disks.

The growth rate is in our simulations relatively slow, mainly due to a low mass transfer-efficiency, the high turbulence kicking the small particles away from the midplane where the boulders are concentrated, and the low dust densities already at 3 AU. This means that it might be necessary to form planetesimals in regions with enhanced densities such as in pressure bumps, where it can be accelerated, and also that there is a need for the radial drift to be prevented over long timescales. We find in this work that planetesimals formed in this way would be small, between 100 meters to some kilometers in size, which is smaller than what was generally predicted by \cite{2010ApJ...724.1153X}, where a larger mass transfer-efficiency was assumed.

This mode of growth is however able to function also in high turbulence regions with relative velocities reaching 100 m s$^{-1}$. This is because the turbulence simultaneously suppresses the size of the bouncing particles, which are therefore more likely to cause mass transfer rather than erosion of the boulders during impacts. Another important effect we have found is that a very small number of seeds are necessary to initiate sweep-up. As the large particles collide with each other, they will create a number of intermediate-sized particles that are also able to sweep-up the small particles, causing the population of large particles to increase not only in total mass, but also in number.

Exactly how these seeds would be introduced into the system needs to be explored further. We have shown that thanks to a combination of more efficient sticking found in recent laboratory experiments and our ability to numerically resolve very small numbers of particles, the bouncing barrier can be overcome or circumvented. If the dust size-distribution is wide enough, a small number of lucky particles might grow over the barrier via hit-and-stick collisions with much smaller particles. This shows that the initial conditions of planet formation might be very important for how the dust growth proceeds. Another option, which we intend to investigate further, is the effect of adding a velocity distribution for each particle size, which would make it possible for some lucky particles to always experience low-velocity collisions and thus grow to large enough sizes.

This work shows where the focus of future laboratory experiments should lie. It is clear that in order for collisional growth of larger particles to be possible, it has to occur between particles of very different sizes. However, very few experiments have been performed to quantify the amount of erosion or mass transfer for varying projectile sizes, porosities, velocities and impact angles, and the understanding of the physical process is still not clear. The maximum size a projectile can have to cause growth of the target due to mass transfer determines whether the small bouncing particles cause erosion or growth of the large boulders. It is also necessary in the lab to determine the smallest projectile size that still leads to growth. Laboratory experiments show that monomer impacts lead to erosion, but whether erosion still occurs at 10 or 100 $\mu m$ is not known. In order to determine in which parameter space fragmentation with mass transfer occurs, more experiments need to be performed.

With an implicit scheme, this code runs fast in terms of execution time, and should be suitable for global disk simulations, something which is not possible with a Monte Carlo approach. In future work, we intend to implement the new collision model in the global dust evolution code of \cite{2010A&A...513A..79B}. This will make it possible to naturally study the number of seeds that can be formed in the disk and how they migrate inwards to other parts of the disk, and also to see how this more sophisticated collision model affects the dust-sizes further out in the disk at 80-100 AU where comparison with observations is possible.


\begin{acknowledgements}
We like to thank Andras Zsom, Chris Ormel, Satoshi Okuzumi, Stefan Kothe, Jens Teiser and Gerhard Wurm for many valuable discussions. F.W. was funded by the Deutsche Forschungsgemeinschaft within the Forschergruppe 759 "The Formation of Planets: The Critical First Growth Phase". C.G. acknowledges financial support from the Japan Society for the Promotion of Science (JSPS). We also like to thank the referee, Stuart Weidenschilling, for his fast and insightful comments that helped improve the paper.
\end{acknowledgements}
\bibliographystyle{aa}
\bibliography{refs}

\begin{thebibliography}{69}
\expandafter\ifx\csname natexlab\endcsname\relax\def\natexlab#1{#1}\fi

\bibitem[{Barge \& Sommeria(1995)}]{1995A&A...295L...1B}
Barge, P. \& Sommeria, J. 1995, A{\&}A, 295, L1

\bibitem[{Beitz {et~al.}(2011)Beitz, G{\"u}ttler, Blum, Meisner, Teiser, \&
  Wurm}]{2011ApJ...736...34B}
Beitz, E., G{\"u}ttler, C., Blum, J., {et~al.} 2011, ApJ, 736, 34

\bibitem[{Beitz {et~al.}(2012)Beitz, G{\"u}ttler, Weidling, \&
  Blum}]{2011arXiv1105.3897B}
Beitz, E., G{\"u}ttler, C., Weidling, R., \& Blum, J. 2012, Icarus, in press

\bibitem[{Benz \& Asphaug(1999)}]{1999Icar..142....5B}
Benz, W. \& Asphaug, E. 1999, Icarus, 142, 5

\bibitem[{Birnstiel {et~al.}(2010)Birnstiel, Dullemond, \&
  Brauer}]{2010A&A...513A..79B}
Birnstiel, T., Dullemond, C.~P., \& Brauer, F. 2010, A{\&}A, 513, 79

\bibitem[{Blum \& M{\"u}nch(1993)}]{1993Icar..106..151B}
Blum, J. \& M{\"u}nch, M. 1993, Icarus, 106, 151

\bibitem[{Blum \& Wurm(2000)}]{2000Icar..143..138B}
Blum, J. \& Wurm, G. 2000, Icarus, 143, 138

\bibitem[{Blum \& Wurm(2008)}]{2008ARA&A..46...21B}
Blum, J. \& Wurm, G. 2008, ARA{\&}A, 46, 21

\bibitem[{Brauer {et~al.}(2008)Brauer, Dullemond, \&
  Henning}]{2008A&A...480..859B}
Brauer, F., Dullemond, C.~P., \& Henning, T. 2008, A{\&}A, 480, 859

\bibitem[{Brauer {et~al.}(2007)Brauer, Dullemond, Johansen, Henning, Klahr, \&
  Natta}]{2007A&A...469.1169B}
Brauer, F., Dullemond, C.~P., Johansen, A., {et~al.} 2007, A{\&}A, 469, 1169

\bibitem[{Ciesla(2009)}]{2009Icar..200..655C}
Ciesla, F.~J. 2009, Icarus, 200, 655

\bibitem[{Cuzzi {et~al.}(2001)Cuzzi, Hogan, Paque, \&
  Dobrovolskis}]{2001ApJ...546..496C}
Cuzzi, J.~N., Hogan, R.~C., Paque, J.~M., \& Dobrovolskis, A.~R. 2001, ApJ,
  546, 496

\bibitem[{Desch(2007)}]{2007ApJ...671..878D}
Desch, S.~J. 2007, ApJ, 671, 878

\bibitem[{Dominik \& Tielens(1997)}]{1997ApJ...480..647D}
Dominik, C. \& Tielens, A. G. G.~M. 1997, ApJ, 480, 647

\bibitem[{Dullemond \& Dominik(2005)}]{2005A&A...434..971D}
Dullemond, C.~P. \& Dominik, C. 2005, A{\&}A, 434, 971

\bibitem[{Geretshauser {et~al.}(2011)Geretshauser, Meru, Speith, \&
  Kley}]{2011A&A...531A.166G}
Geretshauser, R.~J., Meru, F., Speith, R., \& Kley, W. 2011, A{\&}A, 531, 166

\bibitem[{Geretshauser {et~al.}(2010)Geretshauser, Speith, G{\"u}ttler, Krause,
  \& Blum}]{2010A&A...513A..58G}
Geretshauser, R.~J., Speith, R., G{\"u}ttler, C., Krause, M., \& Blum, J. 2010,
  A{\&}A, 513, 58

\bibitem[{G{\"u}ttler {et~al.}(2010)G{\"u}ttler, Blum, Zsom, Ormel, \&
  Dullemond}]{2010A&A...513A..56G}
G{\"u}ttler, C., Blum, J., Zsom, A., Ormel, C.~W., \& Dullemond, C.~P. 2010,
  A{\&}A, 513, 56

\bibitem[{Hayashi {et~al.}(1985)Hayashi, Nakazawa, \&
  Nakagawa}]{1985prpl.conf.1100H}
Hayashi, C., Nakazawa, K., \& Nakagawa, Y. 1985, Protostars and planets II,
  1100

\bibitem[{Hei{\ss}elmann {et~al.}(2007)Hei{\ss}elmann, Blum, \&
  Fraser}]{2007epsc.conf..840H}
Hei{\ss}elmann, D., Blum, J., \& Fraser, H.~J. 2007, European Planetary Science
  Congress 2007, 840

\bibitem[{Ida {et~al.}(2008)Ida, Guillot, \& Morbidelli}]{2008ApJ...686.1292I}
Ida, S., Guillot, T., \& Morbidelli, A. 2008, ApJ, 686, 1292

\bibitem[{Johansen {et~al.}(2008)Johansen, Brauer, Dullemond, Klahr, \&
  Henning}]{2008A&A...486..597J}
Johansen, A., Brauer, F., Dullemond, C.~P., Klahr, H., \& Henning, T. 2008,
  A{\&}A, 486, 597

\bibitem[{Johansen {et~al.}(2011)Johansen, Klahr, \&
  Henning}]{2011A&A...529A..62J}
Johansen, A., Klahr, H., \& Henning, T. 2011, A{\&}A, 529, 62

\bibitem[{Johansen {et~al.}(2007)Johansen, Oishi, Mac~Low, Klahr, Henning, \&
  Youdin}]{2007Natur.448.1022J}
Johansen, A., Oishi, J.~S., Mac~Low, M.-M., {et~al.} 2007, Nature, 448, 1022

\bibitem[{Kelling \& Wurm(2009)}]{2009PhRvL.103u5502K}
Kelling, T. \& Wurm, G. 2009, Phys. Rev. Lett., 103, 215502

\bibitem[{Klahr \& Henning(1997)}]{1997Icar..128..213K}
Klahr, H.~H. \& Henning, T. 1997, Icarus, 128, 213

\bibitem[{Kothe {et~al.}(2010)Kothe, G{\"u}ttler, \&
  Blum}]{2010ApJ...725.1242K}
Kothe, S., G{\"u}ttler, C., \& Blum, J. 2010, ApJ, 725, 1242

\bibitem[{Lammel(2008)}]{Lammel:mLrgVib9}
Lammel, C. 2008, Bachelor's thesis, Technische Universit\"at Carolo Wilhelmina
  Braunschweig

\bibitem[{Langkowski {et~al.}(2008)Langkowski, Teiser, \&
  Blum}]{2008ApJ...675..764L}
Langkowski, D., Teiser, J., \& Blum, J. 2008, ApJ, 675, 764

\bibitem[{Leinhardt \& Stewart(2012)}]{2012ApJ...745...79L}
Leinhardt, Z.~M. \& Stewart, S.~T. 2012, ApJ, 745, 79

\bibitem[{Lynden-Bell \& Pringle(1974)}]{1974MNRAS.168..603L}
Lynden-Bell, D. \& Pringle, J.~E. 1974, MNRAS, 168, 603

\bibitem[{Mizuno(1980)}]{1980PThPh..64..544M}
Mizuno, H. 1980, Progress of Theoretical Physics, 64, 544

\bibitem[{Nakagawa {et~al.}(1981)Nakagawa, Nakazawa, \&
  Hayashi}]{1981Icar...45..517N}
Nakagawa, Y., Nakazawa, K., \& Hayashi, C. 1981, Icarus, 45, 517

\bibitem[{Nakagawa {et~al.}(1986)Nakagawa, Sekiya, \&
  Hayashi}]{1986Icar...67..375N}
Nakagawa, Y., Sekiya, M., \& Hayashi, C. 1986, Icarus, 67, 375

\bibitem[{Okuzumi \& Hirose(2011)}]{2011arXiv1108.4892O}
Okuzumi, S. \& Hirose, S. 2011, arXiv, astro-ph.EP, 4892

\bibitem[{Okuzumi {et~al.}(2009)Okuzumi, Tanaka, \&
  Sakagami}]{2009ApJ...707.1247O}
Okuzumi, S., Tanaka, H., \& Sakagami, M.-a. 2009, ApJ, 707, 1247

\bibitem[{Okuzumi {et~al.}(2011{\natexlab{a}})Okuzumi, Tanaka, Takeuchi, \&
  Sakagami}]{2011ApJ...731...95O}
Okuzumi, S., Tanaka, H., Takeuchi, T., \& Sakagami, M.-a. 2011{\natexlab{a}},
  ApJ, 731, 95

\bibitem[{Okuzumi {et~al.}(2011{\natexlab{b}})Okuzumi, Tanaka, Takeuchi, \&
  Sakagami}]{2011ApJ...731...96O}
Okuzumi, S., Tanaka, H., Takeuchi, T., \& Sakagami, M.-a. 2011{\natexlab{b}},
  ApJ, 731, 96

\bibitem[{Ormel \& Cuzzi(2007)}]{2007A&A...466..413O}
Ormel, C.~W. \& Cuzzi, J.~N. 2007, A{\&}A, 466, 413

\bibitem[{Ormel \& Spaans(2008)}]{2008ApJ...684.1291O}
Ormel, C.~W. \& Spaans, M. 2008, ApJ, 684, 1291

\bibitem[{Paraskov {et~al.}(2007)Paraskov, Wurm, \&
  Krauss}]{2007Icar..191..779P}
Paraskov, G.~B., Wurm, G., \& Krauss, O. 2007, Icarus, 191, 779

\bibitem[{Paszun \& Dominik(2009)}]{2009A&A...507.1023P}
Paszun, D. \& Dominik, C. 2009, A{\&}A, 507, 1023

\bibitem[{Pinilla {et~al.}(2011)Pinilla, Birnstiel, Ricci, Dullemond, Uribe,
  Testi, \& Natta}]{2011arXiv1112.2349P}
Pinilla, P., Birnstiel, T., Ricci, L., {et~al.} 2011, arXiv, 1112, 2349

\bibitem[{Pollack {et~al.}(1996)Pollack, Hubickyj, Bodenheimer, Lissauer,
  Podolak, \& Greenzweig}]{1996Icar..124...62P}
Pollack, J.~B., Hubickyj, O., Bodenheimer, P., {et~al.} 1996, Icarus, 124, 62

\bibitem[{Schr{\"a}pler \& Blum(2011)}]{2011ApJ...734..108S}
Schr{\"a}pler, R. \& Blum, J. 2011, ApJ, 734, 108

\bibitem[{Sekiya \& Takeda(2003)}]{2003EP&S...55..263S}
Sekiya, M. \& Takeda, H. 2003, Earth, 55, 263

\bibitem[{Sekiya \& Takeda(2005)}]{2005Icar..176..220S}
Sekiya, M. \& Takeda, H. 2005, Icarus, 176, 220

\bibitem[{Stewart \& Leinhardt(2011)}]{2011arXiv1109.4588S}
Stewart, S.~T. \& Leinhardt, Z.~M. 2011, arXiv, 1109, 4588

\bibitem[{Tanaka {et~al.}(2005)Tanaka, Himeno, \& Ida}]{2005ApJ...625..414T}
Tanaka, H., Himeno, Y., \& Ida, S. 2005, ApJ, 625, 414

\bibitem[{Teiser {et~al.}(2011)Teiser, K{\"u}pper, \&
  Wurm}]{2011arXiv1108.0785T}
Teiser, J., K{\"u}pper, M., \& Wurm, G. 2011, arXiv, 1108, 785

\bibitem[{Teiser \& Wurm(2009{\natexlab{a}})}]{2009A&A...505..351T}
Teiser, J. \& Wurm, G. 2009{\natexlab{a}}, A{\&}A, 505, 351

\bibitem[{Teiser \& Wurm(2009{\natexlab{b}})}]{2009MNRAS.393.1584T}
Teiser, J. \& Wurm, G. 2009{\natexlab{b}}, MNRAS, 393, 1584

\bibitem[{Thornton \& Ning(1998)}]{Thornton:1998ve}
Thornton, C. \& Ning, Z. 1998, Powder Technology, 99, 154

\bibitem[{Wada {et~al.}(2007)Wada, Tanaka, Suyama, Kimura, \&
  Yamamoto}]{2007ApJ...661..320W}
Wada, K., Tanaka, H., Suyama, T., Kimura, H., \& Yamamoto, T. 2007, ApJ, 661,
  320

\bibitem[{Wada {et~al.}(2008)Wada, Tanaka, Suyama, Kimura, \&
  Yamamoto}]{2008ApJ...677.1296W}
Wada, K., Tanaka, H., Suyama, T., Kimura, H., \& Yamamoto, T. 2008, ApJ, 677,
  1296

\bibitem[{Wada {et~al.}(2009)Wada, Tanaka, Suyama, Kimura, \&
  Yamamoto}]{2009ApJ...702.1490W}
Wada, K., Tanaka, H., Suyama, T., Kimura, H., \& Yamamoto, T. 2009, ApJ, 702,
  1490

\bibitem[{Wada {et~al.}(2011)Wada, Tanaka, Suyama, Kimura, \&
  Yamamoto}]{2011ApJ...737...36W}
Wada, K., Tanaka, H., Suyama, T., Kimura, H., \& Yamamoto, T. 2011, ApJ, 737,
  36

\bibitem[{Weidenschilling(1977{\natexlab{a}})}]{1977MNRAS.180...57W}
Weidenschilling, S.~J. 1977{\natexlab{a}}, MNRAS, 180, 57

\bibitem[{Weidenschilling(1977{\natexlab{b}})}]{1977Ap&SS..51..153W}
Weidenschilling, S.~J. 1977{\natexlab{b}}, Astrophys. {\&} Space Sci., 51, 153

\bibitem[{Weidenschilling(1980)}]{1980Icar...44..172W}
Weidenschilling, S.~J. 1980, Icarus, 44, 172

\bibitem[{Weidenschilling(1997)}]{1997Icar..127..290W}
Weidenschilling, S.~J. 1997, Icarus, 127, 290

\bibitem[{Weidenschilling \& Cuzzi(1993)}]{1993prpl.conf.1031W}
Weidenschilling, S.~J. \& Cuzzi, J.~N. 1993, In: Protostars and planets III
  (A93-42937 17-90), 1031

\bibitem[{Weidling {et~al.}(2012)Weidling, G{\"u}ttler, \&
  Blum}]{2011arXiv1105.3909W}
Weidling, R., G{\"u}ttler, C., \& Blum, J. 2012, Icarus, in press

\bibitem[{Whipple(1972)}]{Whipple:3gV1B7CM}
Whipple, F. 1972, {in From Plasma to Planet} (New York: Wiley Interscience
  Divison)

\bibitem[{Wurm {et~al.}(2005)Wurm, Paraskov, \& Krauss}]{2005Icar..178..253W}
Wurm, G., Paraskov, G.~B., \& Krauss, O. 2005, Icarus, 178, 253

\bibitem[{Xie {et~al.}(2010)Xie, Payne, Th{\'e}bault, Zhou, \&
  Ge}]{2010ApJ...724.1153X}
Xie, J.-W., Payne, M.~J., Th{\'e}bault, P., Zhou, J.-L., \& Ge, J. 2010, ApJ,
  724, 1153

\bibitem[{Zsom \& Dullemond(2008)}]{2008A&A...489..931Z}
Zsom, A. \& Dullemond, C.~P. 2008, A{\&}A, 489, 931

\bibitem[{Zsom {et~al.}(2011)Zsom, Ormel, Dullemond, \&
  Henning}]{2011A&A...534A..73Z}
Zsom, A., Ormel, C.~W., Dullemond, C.~P., \& Henning, T. 2011, A{\&}A, 534, 73

\bibitem[{Zsom {et~al.}(2010)Zsom, Ormel, G{\"u}ttler, Blum, \&
  Dullemond}]{2010A&A...513A..57Z}
Zsom, A., Ormel, C.~W., G{\"u}ttler, C., Blum, J., \& Dullemond, C.~P. 2010,
  A{\&}A, 513, 57

\end{thebibliography}

\makeatletter
\if@referee
\processdelayedfloats
\pagestyle{plain}
\fi
\makeatother
\end{document}